%% file: paper.tex
\documentclass[conference,compsoc]{IEEEtran}

\usepackage[utf8]{inputenc}
\usepackage[T1]{fontenc}

\usepackage[table]{xcolor}
\usepackage{graphicx}
\usepackage{caption}
\usepackage{subcaption}
\usepackage{stfloats}
\usepackage{float}
\usepackage{csquotes}
\usepackage{subcaption}
\usepackage{amsthm}

\usepackage[weather]{ifsym}
\usepackage[colorlinks=true, urlcolor=black, linkcolor=black]{hyperref}
\usepackage{paralist}
\renewenvironment{itemize}{\begin{compactitem}}{\end{compactitem}}
\setdefaultenum{(1.)}{}{}{}
\setdefaultitem{$\blacktriangleright$}{}{}{}
\usepackage{listings}
\renewcommand*{\paragraph}[1]{\vspace{2mm}\noindent\textbf{#1.}}

\usepackage{setspace}
\usepackage{array}
\newcolumntype{x}[1]{>{\centering\let\newline\\\arraybackslash\hspace{0pt}}p{#1}}
\usepackage{threeparttable}

\lstdefinestyle{base}{
  emptylines=1,
  breaklines=true,
  basicstyle=\ttfamily\scriptsize\color{black},
  moredelim=**[is][\color{red}]{@}{@},
}

\input{stuff/math}

\input{stuff/comment}
\input{stuff/openid_macros}
\input{stuff/glossaries}
\input{stuff/lstlistingstyles}

\author{
\IEEEauthorblockN{Christian Mainka}
	\IEEEauthorblockA{%
		Horst Görtz Institute for IT-Security\\
		Ruhr University Bochum\\
		\texttt{Christian.Mainka@rub.de}}
\and
\IEEEauthorblockN{Vladislav Mladenov}
	\IEEEauthorblockA{%
		Horst Görtz Institute for IT-Security\\
		Ruhr University Bochum\\
		\texttt{Vladislav.Mladenov@rub.de}}
\and
\IEEEauthorblockN{Jörg Schwenk}
	\IEEEauthorblockA{%
	Horst Görtz Institute for IT-Security\\
	Ruhr University Bochum\\
	\texttt{Joerg.Schwenk@rub.de}}
}

\begin{document}

\title{\Large \bf On the security of modern Single Sign-On Protocols -- \\ Second-Order Vulnerabilities in OpenID Connect}

\maketitle

\input{sections/abstract}
\glsresetall{}
\input{sections/introduction}

\input{sections/basics}

\input{sections/attackerModel}
\input{sections/evaluationGap}
\input{sections/secondorder}
\input{sections/attacks}
\input{sections/implementation}

\input{sections/countermeasures}

\input{sections/relatedWork}

\input{sections/conclusion}

{\footnotesize \bibliographystyle{splncs03}
\bibliography{literature}

\end{document}

%% file: stuff/math.tex
\usepackage{amsmath,amssymb,wasysym}
\DeclareMathAlphabet{\mathcal}{OMS}{cmsy}{m}{n}

%% file: stuff/comment.tex
\usepackage{comment}
\specialcomment{overview}{\textbf{Overview:}}{}

%% file: stuff/glossaries.tex
\usepackage[xindy]{glossaries}

\glossarystyle{indexgroup}
\makeglossaries

\newacronym[]{trc}{TRC}{Token Recipient Confusion}
\newacronym[]{kc}{KC}{Key Confusion}
\newacronym[]{ids}{IDS}{ID Spoofing}
\newacronym[]{ds}{DS}{Discovery Spoofing}
\newacronym[%
description={\glsentrylong{aaas}; A SOA concept which provides an authentication service as a service},
]
{aaas}{AaaS}{authentication as a service}
\newacronym[%
description={\glsentrylong{ax}; OpenID extension for exchanging identity information between endpoints~\cite{ax}},
]
{ax}{Ax}{Attribute Exchange}
\newacronym[%
    description={\glsentrylong{sreg}; OpenID extension for exchanging identity information between endpoints~\cite{sreg}},
]
{sreg}{SReg}{Simple Registration}
\newacronym[%
description={\glsentrylong{cms}; A software which helps to create an manage content collaboratory. CMSs are often web applications},
]
{cms}{CMS}{content-management system}
\newacronym[%
description={\glsentrylong{idp}; An entity which is responsible for creating tokens that will be used to authenticate a user to an SP},
]
{idp}{IdP}{Identity Provider}
\newacronym[%
description={\glsentrylong{dos}; The unavailability of a service, which should be available},
]
{dos}{DoS}{Denial-of-Service}
\newacronym[%
description={\glsentrylong{gui}; A software component which allows a human to interact with machine by using symbols},
]
{gui}{GUI}{Graphical User Interface}
\newacronym[%
description={\glsentrylong{jaxb}; An interface for Java allowing to bind XML Data to Java objects. See \url{https://jaxb.java.net/}},
]
{jaxb}{JAXB}{Java Architecture for XML Binding}
\newacronym[%
description={\glsentrylong{mitm}; An attack technique in which the attacker virtually or physically sits between to users, allowed to listen and manipulate their exchanged messages},
]
{mitm}{MitM}{Man-in-the-Middle}
\newacronym[
    description={Single Sign-On is a property of access controll, which allows a user to log in once and request access to several systems},
]{sso}{SSO}{Single Sign-On}
\newacronym[%
description={\glsentrylong{soa}; Abstract model of software architecture},
]
{soa}{SOA}{Service Oriented Architecture}
\newacronym[%
description={\glsentrylong{sp}; Also known as \emph{Relying Party}, entity that offers a service which the user wants to use},
]
{sp}{SP}{Service Provider}
\newacronym[%
description={A user agent is software (a software agent) that is acting on behalf of a user (Wikipedia)}
]
{ua}{UA}{user agent}
\newacronym[%
description={\glsentrylong{w3c}; Organization for standards in the World Wide Web},
]
{w3c}{W3C}{World Wide Web Consortium}
\newacronym[%
description={\glsentrylong{xml}; Textual data format to encode documents, commonly used for message exchange~\cite{xml}},
]
{xml}{XML}{eXtended Markup Language}
\newacronym[%
description={\glsentrylong{xsw}; Technique for attacking signed XML documents~\cite{McIntosh2005}},
]
{xsw}{XSW}{XML Signature Wrapping}

\newglossaryentry{dhke}{%
    name={Diffie-Hellman key exchange},
    description={Specific method for exchanging key material},
}
\newglossaryentry{asso}{%
    name={association},
    description={An association between the SP and the OpenID IdP establishes a shared secret between them~\cite[Section 8]{openid20}},
}
\newglossaryentry{csrf}{%
    name={CSRF},
    description={Cross-Site-Request-Forgery; Attack technique which tries to use web application APIs by a victim without his knowledge},
}
\newglossaryentry{dirver}{%
    name={direct verification},
    description={OpenID signature verfication performed by the OpenID IdP itself, enforced by the SP~\cite[Section 11.4.2]{openid20}},
}
\newglossaryentry{junit}{%
    name={JUnit},
    description={JUnit is framework for testing Java programs},
}
\newglossaryentry{testng}{%
    name={TestNG},
    description={TestNG is framework for testing Java programs},
}
\newglossaryentry{mysql}{%
    name={MySQL},
    description={MySQL is a widely used \glsentrytext{opensource} relational database management system},
}
\newglossaryentry{oauth}{%
    name={OAuth},
    description={OAuth is an open protocol which standardized API-Authoring for desktop-, web-, and mobile-applications~\cite{oauth}},
}
\newglossaryentry{oid}{%
    name={OpenID},
    description={OpenID is decentralized authentication system for web-based SPs~\cite{openid20}},
}
\newglossaryentry{oid4j}{%
    name={OpenID4Java},
    description={OpenID4Java is library which easily allows to implement an OpenID consumber as well as an OpenID IdP ~\cite{openid4java}},
}
\newglossaryentry{oida}{%
    name={OpenID Attacker},
    description={OpenID Attacker is the tool developed in this thesis for attacking the OpenID specification},
}
\newglossaryentry{opensource}{%
    name={open source},
    description={Work which are enforced by license to have source available for the public},
}
\newglossaryentry{oc}{
    name={ownCloud},
    description={OwnCloud is a free and \glsentrytext{opensource}, \glsentrytext{php} and \glsentrytext{mysql} based web application which allows data synchronization and cloud storage~\cite{owncloud}.}
}
\newglossaryentry{pentest}{%
    name={penetration test},
    user1={Penetration Test},
    user2={Penetration Tests},
    user3={penetration testing},
    user4={Penetration Testing},
    user5={penetration tester},
    user6={penetration testers},
    description={Method for evaluating security on computer systems},
}
\newglossaryentry{php}{%
    name={PHP},
    description={PHP is a popular server-side scripting language and widespread in the area of web development},
}
\newglossaryentry{saml}{%
    name={SAML},
    description={Security Assertion Markup Language~\cite{saml}},
}
\newglossaryentry{sqlinjection}{
    name={SQL-Injection},
    description={An attack technique which tries to inject and execute SQL statements reasoned in inadequate user input validation~\cite{sqlinjection}},
}
\newglossaryentry{wsattacker}{%
    name={WS-Attacker},
    description={Automatic \glsentrytext{pentest} framework~\cite{wsattacker}},
}
\newglossaryentry{doc}{%
    name={XML document},
    description={General term for a document which uses \glsentrytext{xml} as description language, e.g.\ a \glsentrytext{soap} message},
    parent=xml,
    user1={document},
    user2={documents},
}
\newglossaryentry{wa}{%
    name={web application},
    user1={Web Application},
    user2={Web Applications},
    description={A web application is an application that uses a web browser as a client.}
}
\newglossaryentry{wp}{%
    name={WordPress},
    description={WordPress is a free and \glsentrytext{opensource}, \glsentrytext{php} and \glsentrytext{mysql} based \glsentrytext{cms}~\cite{wordpress}.}
}
\newglossaryentry{xmlsig}{%
    name={XML Signature},
    user1={XML Digital Signature},
    description={also \glsentryuseri{xmlsig}; Standard for creating signatures in \glsentrytext{xml} documents~\cite{xmlsig1, xmlsig2}},
}
\newglossaryentry{xrds}{%
    name={XRDS},
    description={eXtensible Resource Descriptor Sequence is an XML format for describing metadata as a web resource},
}
\newglossaryentry{xss}{%
    name={XSS},
    description={Cross-Site-Scripting is a web application vulnerability which tries to execute an attacker script within the context of the website owner within the user's browser},
}

\newglossaryentry{enduser}{%
    name={End-User},
    description={A user surfing in the web via his User-Agent},
}
\newglossaryentry{http}{%
	name={HTTP},
	description={},
}

\newacronym[longplural=Relying Parties]{rp}{RP}{Relying Party}
\newacronym[]{op}{OP}{OpenID Provider}
\newacronym[]{kdc}{KDC}{Key-Distribution Center}
\newacronym[]{tls}{TLS}{Transport Layer Security}
\newacronym[]{https}{HTTPS}{Hypertext Transfer Protocol Secure / HTTP Over TLS}
\newacronym[]{url}{URL}{Uniform Resource Locator}
\newacronym[]{uri}{URI}{Uniform Resource Identifier}
\newacronym[]{api}{API}{Application Programming Interface}
\newacronym[]{rest}{REST}{Representational State Transfer}
\newacronym[]{json}{JSON}{JavaScript Object Notation}
\newacronym[]{jwt}{JWT}{JSON Web Token}
\newacronym[]{jwks}{JWKS}{JSON Web Key Set}
\newacronym[]{jws}{JWS}{JSON Web Signature}
\newacronym[]{jwe}{JWE}{JSON Web Encryption}
\newacronym[]{ic}{IC}{Issuer Confusion}
\newacronym[]{sm}{SM}{Signature Manipulation}
\newacronym[]{scs}{SCS}{Sub Claim Spoofing}
\newacronym[]{rum}{RUM}{Redirect URI Manipulation}
\newacronym[]{ietf}{IETF}{Internet Engineering Task Force}
\newacronym[]{arm}{aRM}{Attacker Related Message}
\newacronym[]{brm}{BRM}{Browser Relayed Message}
\newacronym[]{ttp}{TTP}{Trusted Third Party}
\newacronym[]{ta}{TA}{tainted address}
\newacronym[]{ssrf}{SSRF}{Server Side Request Forgery}

\newglossaryentry{oidcp}{%
    name={OpenID Connect protocol},
    description={},
}

\newglossaryentry{oidc}{%
    name={OpenID Connect},
    description={},
}

\newglossaryentry{mep}{name=Malicious Endpoints,description=}
\newglossaryentry{mds}{name=malicious Discovery service,description=}
\newglossaryentry{redirect_uri}{name=\texttt{redirect\_uri},description=}
\newglossaryentry{client_id}{name=\texttt{client\_id},description=}
\newglossaryentry{scope}{name=\texttt{scope},description=}
\newglossaryentry{response_type}{name=\texttt{response\_type},description=}
\newglossaryentry{code}{name=\texttt{code},description=}
\newglossaryentry{state}{name=\texttt{state},description=}
\newglossaryentry{grant_type}{name=\texttt{grant\_type},description=}
\newglossaryentry{orization_code}{name=\texttt{authorization\_code},description=}
\newglossaryentry{client_secret}{name=\texttt{client\_\-secret},description=}
\newglossaryentry{token_type}{name=\texttt{token\_type},description=}
\newglossaryentry{expires_in}{name=\texttt{expires\_in},description=}
\newglossaryentry{refresh_token}{name=\texttt{refresh\_token},description=}
\newglossaryentry{iss}{name=\texttt{issuer},description=}
\newglossaryentry{sub}{name=\texttt{subject},description=}
\newglossaryentry{aud}{name=\texttt{audience},description=}
\newglossaryentry{iat}{name=\texttt{timestamp},description=}
\newglossaryentry{exp}{name=\texttt{expired},description=}
\newglossaryentry{nonce}{name=\texttt{nonce},description=}
\newglossaryentry{id_token}{name=\texttt{id\_token},description=}
\newglossaryentry{access_token}{name=\texttt{access\_token},description=}
\newglossaryentry{token}{name=\texttt{token},description=}
\newglossaryentry{resource}{name=\texttt{resource},description=}
\newglossaryentry{rel}{name=\texttt{rel},description=}
\newglossaryentry{href}{name=\texttt{href},description=}
\newglossaryentry{issuer}{name=\texttt{issuer},description=}
\newglossaryentry{authorization_endpoint}{name=\texttt{authorization\_endpoint},description=}
\newglossaryentry{token_endpoint}{name=\texttt{token\_endpoint},description=}
\newglossaryentry{jwks_uri}{name=\texttt{jwks\_uri},description=}
\newglossaryentry{response_types_supported}{name=\texttt{response\_types\_supported},description=}
\newglossaryentry{subject_types_supported}{name=\texttt{subject\_types\_supported},description=}
\newglossaryentry{id_token_signing_alg_values_supported}{name=\texttt{id\_token\_\-signing\_alg\_values\_supported},description=}
\newglossaryentry{application_type}{name=\texttt{application\_type},description=}
\newglossaryentry{redirect_uris}{name=\texttt{redirect\_uris},description=}
\newglossaryentry{azp}{name=\texttt{azp},description=}
\newglossaryentry{claims}{name=\texttt{claims},description=}
\newglossaryentry{request}{name=\texttt{request},description=}
\newglossaryentry{request_uri}{name=\texttt{request\_uri},description=}
\newglossaryentry{auth_time}{name=\texttt{auth\_time},description=}
\newglossaryentry{acr}{name=\texttt{acr},description=}
\newglossaryentry{client_secret_basic}{name=\texttt{client\_secret\_basic},description=}
\newglossaryentry{client_secret_post}{name=\texttt{client\_secret\_post},description=}
\newglossaryentry{client_secret_jwt}{name=\texttt{client\_secret\_jwt},description=}
\newglossaryentry{private_key_jwt}{name=\texttt{private\_key\_jwt},description=}
\newglossaryentry{claims_parameter_supported}{name=\texttt{claims\_parameter\_supported},description=}
\newglossaryentry{request_parameter_supported}{name=\texttt{request\_parameter\_supported},description=}
\newglossaryentry{at_hash}{name=\texttt{at\_hash},description=}
\newglossaryentry{c_hash}{name=\texttt{c\_hash},description=}

%% file: sections/abstract.tex
\subsection*{Abstract}
\label{sec:abstract}

\Gls{oauth} is the new de facto standard for delegating authorization in the web.
An important limitation of \gls{oauth} is the fact that it was designed for authorization and not for authentication. 
The usage of \gls{oauth} for authentication thus leads to serious vulnerabilities as shown by Zhou et. al. in \cite{ssoscan} and Chen et. al. in \cite{demystifiedOAuthCCS14}.
\gls{oidc} was created on top of \gls{oauth} to fill this gap by providing federated identity management and user authentication. %
\gls{oidc} was standardized in February 2014, but leading companies like Google, Microsoft, AOL and PayPal are already using it in their web applications~\cite{oidc_google,oidc_paypal,oidc_microsoft,oidc_leadership}.

In this paper we describe the \gls{oidcp} and provide the first in-depth analysis of one of the key features of \gls{oidc}: the \emph{Discovery} and the \emph{Dynamic Registration} extensions.
We present a new class of attacks on \gls{oidc} that belong to the category of \emph{second-order vulnerabilities}. 
These attacks consist of two phases: First, the injection payload is stored by the legitimate application. Later on, this payload is used in a security-critical operation.

Our new class of attacks -- called \emph{Malicious Endpoints attacks} -- exploits the \gls{oidc} extensions \emph{Discovery} and \emph{Dynamic Registration}. These attacks break user authentication, compromise user privacy, and enable \gls{ssrf}, client-side code injection, and \gls{dos}.
As a result, the security of the \gls{oidc} protocol cannot be guaranteed when these extensions are enabled in their present form.

We contacted the authors of the \gls{oidc} and \gls{oauth} specifications.
They acknowledged our Malicious Endpoint attacks and recognized the need to improve the specification~\cite{discoverySecurityConsiderations}.
We are currently involved in the discussion regarding the mitigation of the existing issues and an extension to the \gls{oauth} specification.

%% file: sections/introduction.tex
\section{Introduction}
\label{sec:intro}

\paragraph{\gls{sso}}
\gls{sso} protocols like SAML, \gls{oid} or \gls{oidc} replace multiple manual authentications at different \glspl{sp} with a single manual authentication at an \gls{idp}, and multiple REST-based messages invisible to the \gls{enduser}.
An \gls{idp} manages identities of multiple \gls{enduser}s, provides specific authentication mechanisms (e.g., username/password or 2-factor), and creates authentication tokens about authenticated \gls{enduser}s.
These authentication tokens are consumed by an \gls{sp} granting or denying access to the \gls{enduser} in dependence of the token verification. 

\paragraph{Security of \gls{sso}}
Many known attacks on \gls{sso} systems only tamper with one protocol step and achieve the desired results in the following step.
For example, replay-attacks, or attacks manipulating the token directly or sending it to a different \gls{sp}~\cite{grossSAML,saas_ccsw14,SoMaScKaJeSAML12,microsoft} -- they all achieve their attack goals in a single protocol request/response pair.
We thus classify these vulnerabilities as {\em first-order} vulnerabilities, since they can be detected by only checking a single control flow.
Modern analyzing tools like SSOScan~\cite{ssoscan}, AuthScan~\cite{authscan} and InteGuard~\cite{integuard} are able to detect such first-order vulnerabilities but are limited to one protocol or cover only a small subset of existing attacks.

A more general approach is the automated analysis of \gls{sso} protocols, which remains a future challenge: Only the relatively simple flows of the SAML \gls{sso} protocol have been analyzed with protocol analyzers \cite{grossSAML}. Sun et al. in 2012 ~\cite{journals/compsec/SunHB12} and Fett et al. in 2014 ~\cite{FettKuestersSchmitz-SP-2014} proposed a formal model for analyzing \gls{oid} and Browser ID, but admit that the protocols are far too complex to be analyzed automatically.

Summarized, previous work concentrated on first-order vulnerabilities in \gls{saml} \cite{grossSAML,saas_ccsw14,SoMaScKaJeSAML12}, \gls{oid} \cite{sessionSwap,tsyr,journals/compsec/SunHB12}, \gls{oauth}~\cite{demystifiedOAuthCCS14,sun2012devil} and Facebook Connect~\cite{miculan2011formal,ssoscan}, but the \gls{oidcp} and especially its extensions have not been investigated so far.

\paragraph{\gls{oidc}}
\gls{oidc} is a new \gls{sso} protocol released in February 2014. It is the successor of \gls{oid} and it is based on \gls{oauth}, but uses several ideas from \gls{oid}.

The key feature of \gls{oid} --- the dynamic and fully automatic \emph{open} trust establishment between \gls{idp} and \gls{sp} --- is also present in the \gls{oidcp} by means of the \emph{Discovery} and \emph{Dynamic Registration} extensions.
\emph{Open} in the context of \gls{sso} means that users can be logged into an \gls{sp} even if the user's \gls{idp} is not known to the \gls{sp} beforehand.
A user can simply submit its identity on the \gls{sp}, which is usually a URL (e.g., \emph{\url{https://IdP.com/alice}}) or an email address (e.g., \emph{alice@Idp.com}).
Based on this identity the \gls{sp} discovers the responsible \gls{idp} (e.g., \emph{\url{https://IdP.com/}}) and retrieves all information needed for the authentication. %
Afterwards, the \gls{sp} dynamically registers on the discovered \gls{idp} and establishes a trust relationship to be able to (retrieve and) verify the authentication tokens used later on in the \gls{sso} protocol flow.

During \emph{Discovery} and \emph{Dynamic Registration},  \gls{sp} and  \gls{idp} communicate directly with each other (server-to-server communication), so these protocol messages cannot be monitored by the \gls{enduser}. 

\paragraph{Second-Order Vulnerabilities in \gls{oidc}}
Second-order vulnerabilities have been described and detected in the context of web applications \cite{bau2010state,Dahse2014,Olivo2015}.
Speaking of second-order vulnerabilities in \gls{sso} protocols in general we have the same execution scheme:
\begin{inparaenum}
	\item The injection of the attack vectors is allowed by the specification and protocol flow. Thus, no implementation or configuration flaws are required. 
	\item The execution of the protocol can proceed as usual without any incidents. 
	\item The attack vectors are loaded and lead to successful execution of the attack.
\end{inparaenum}

Analyzing and detecting second-order vulnerabilities in distributed systems like \gls{sso} is more complex than in web applications, because they are including multiple phases, plethora messages, parameters and participants.
This makes detection significantly more complex. 
We are not aware of any previous work and any automated security tools capable to detect such vulnerabilities.

\paragraph{Malicious Endpoint Attacks}
The concept of our \emph{Malicious Endpoint attacks} abuses a weakness in the Discovery and Dynamic Registration extensions of the \gls{oidcp} to initially store the payload on the \gls{sp} and execute it in another step.
The main reason for this is that an \gls{sp} can be forced to start a Discovery on an attacker-controlled webserver, which returns attacker-chosen information.
This information contains URL parameters that can be used for different threats. 
For instance we could use them to start \gls{ssrf} targeting the internal network behind the \gls{sp}, execute \gls{dos} attacks by forcing the \gls{sp} to download huge data files, start code injection attacks, and even broke the user authentication on the \gls{sp} --- we were able to steal the user's \gls{sso} token.

\paragraph{Our Contribution} \\
\begin{itemize}
 \item We are the first providing an in-depth security analysis of the \gls{oidc} features \emph{Discovery} and \emph{Dynamic Registration}.
 \item We identified serious second-order vulnerabilities and developed a new class of attacks called \emph{Malicious Endpoints attacks}, which exploit a lack in the \gls{oidc} specification resulting in \gls{ssrf}, \gls{dos}, and authentication flaws.
 \item We propose countermeasures to prevent our attacks, and discuss their respective advantages and disadvantages.
       The integration of our countermeasures are currently discussed with the authors of the specification.
 \item We provide a public available online website that can be used by \glspl{sp} for verifying the security against our Malicious Endpoint attacks.\footnote{The website does not provide any tests against \gls{dos} and \gls{ssrf} attacks in order to avoid misuse. }
\end{itemize}
\vspace{2mm}

\noindent Our results show that protocol extensions must be designed with extreme care, and their security implications have to be discussed thoroughly.
Otherwise, they can lead to serious attacks with critical impact, even in secure standards.

The Discovery and Dynamic Registration are optional extensions. %
Four libraries are officially certified to specific conformance profiles and interoperability~\cite{oidcCeritification}.
We successfully verified our attacks against two of them:  MITREidConnect and phpOIDC.
However, it must be considered that our Malicious Endpoints attack targets on the \gls{oidc} specification itself and not on a specific implementation.
Thus, any implementation using the \emph{Discovery} and \emph{Dynamic Registration} extensions is vulnerable against the class of Malicious Endpoints attacks.

%% file: sections/basics.tex
\section{Modern SSO Protocols} 
\label{sec:basics}

Since their establishment in the early 1980s, protocols like Kerberos (first officially published in 1987 as Version 4~\cite{miller1987kerberos}) and the corresponding concepts of delegated authentication and authorization using \glspl{ttp} have been constantly developed and refined into modern \gls{sso} protocols.
These protocols aim at being compliant with the requirements of the modern and flexible Internet infrastructure.
Mainly, modern \gls{sso} protocols strive to achieve the following goals:

(1) \textbf{Decentralization} --
\gls{sso} appears to be centralized embracing only a very small set of fixed \glspl{ttp}.
The most widely known of these \glspl{ttp} are Facebook and Google.
However, exporting data and outsourcing infrastructure to companies like Facebook and Google can include certain security risks and trust issues.

Luckily, modern protocols like \gls{oauth}, \gls{saml}, BrowserID, \gls{oid} and \gls{oidc} are designed and specified to set up custom \glspl{ttp}, which act independently from each other.
This enables companies to set up their own \glspl{ttp} and use these for authentication purposes instead of having to rely on external providers.

(2) \textbf{Trust Establishment} -- Every \gls{sp} has to establish a trust relationship with the \gls{ttp}. 
In order to do this, key material has to be exchanged. 
An important requirement for modern \gls{sso} protocols is that this process occurs with minimal configuration, implementation, and installation effort.
In the best case, the trust establishment should work automatically. %

\subsection{\gls{oidc}}

The \gls{oidcp} efficiently addresses the goals stated above -- it is decentralized and allows automated trust establishment without any configuration effort or user interaction.

\gls{oidc} was designed on basis of the \gls{oauth} framework~\cite{rfc6749oauth} in order to enable the authentication of \textit{End-Users} without changing the \gls{oauth} protocol flow.
Thus, \gls{oauth} capabilities are integrated with the protocol itself~\cite{OpenIDFoundation2014}, providing \gls{oidc} with the capability of delegated authorization.

Additionally, \gls{oidc} also incorporates concepts utilized by another SSO protocol -- OpenID~\cite{oidcVSoid}.
Such concepts are the Discovery and Dynamic Registration of \glspl{op}.
The Discovery process allows an \gls{sp} to automatically discover new \glspl{op} without any configuration and user interaction.
The Dynamic Registration enables the on-the-fly registration and trust establishment between a Client and \gls{op}, also without any user interaction.

A major advantage of \gls{oidc} is its integration into existing applications:
\Gls{oidc} was designed to be easily integrated into current \gls{oauth} compliant systems, with only minimal extensions to the already available \gls{oauth} APIs.

\subsection{Roles}

\begin{figure}[!ht]
	\centering
	\includegraphics[width=0.35\textwidth]{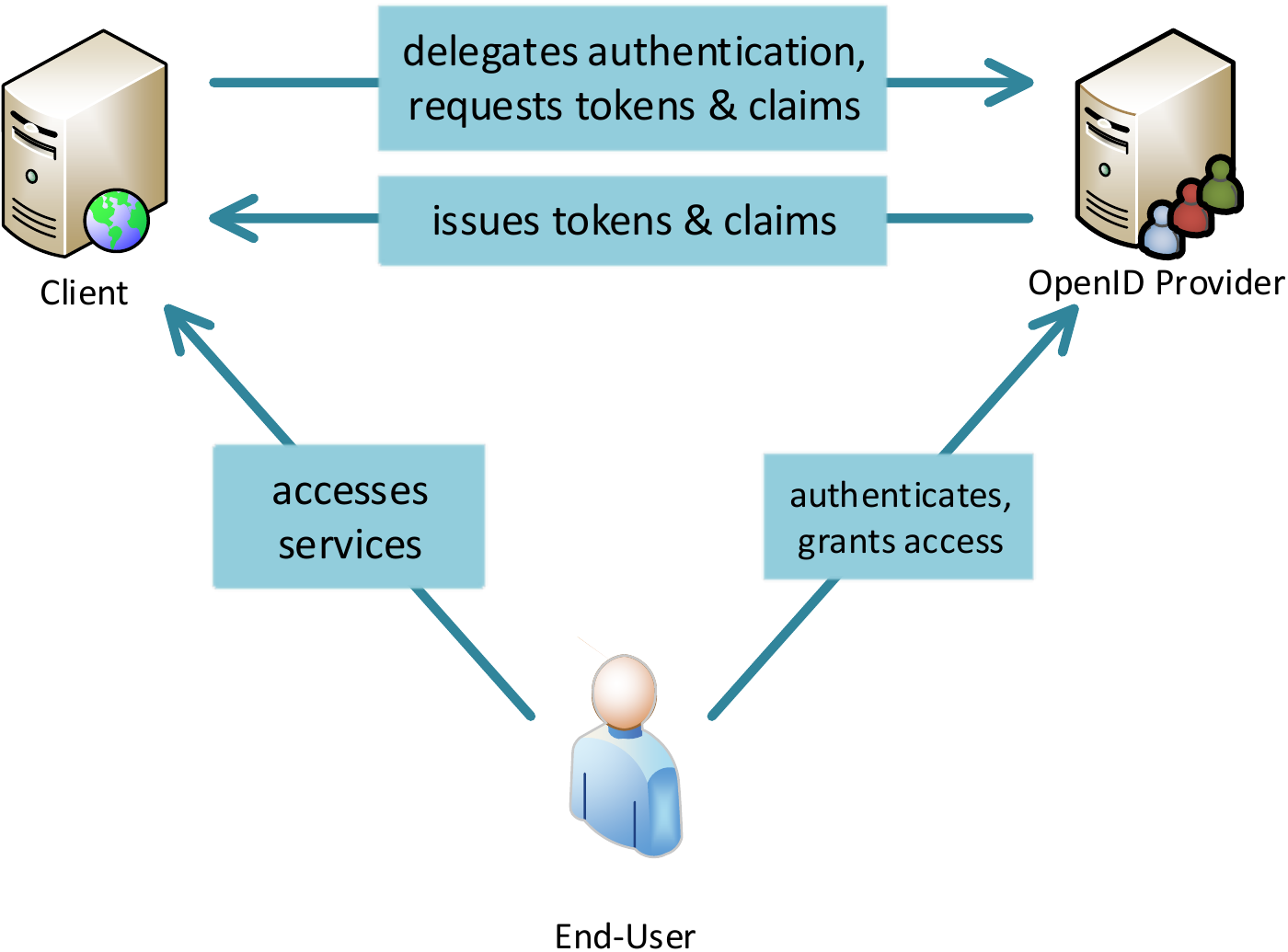}
	\caption[Role Relationship within the \gls{oidcp}]{Role Relationship within the \gls{oidcp}~\cite[Section 1.3]{OpenIDFoundation2014}}
	\label{fig:openidConnectRoles}
\end{figure}

Within the \gls{oidcp}, three different parties each assuming a different role can be found.
The relationship between the different roles can be seen in \autoref{fig:openidConnectRoles}.

\textbf{The End-User}, represented by his \gls{ua}, wants to access selected services of a Client.
Therefore, he needs to prove his identity to the Client.
Additionally, the End-User has the possibility to authorize the Client to access a specific set of his resources stored on the \gls{op}.

\textbf{The Client} is an application providing a certain service, which requires authentication of an End-User.
This authentication process is delegated to the corresponding \gls{op}.
Therefore, the Client requests an authentication token signed by \gls{op}, which proves the identity of the End-User.
Optionally, the Client can also request authorization to access certain protected resources of the End-User stored on \gls{op}, for example, photos.

Please note that the term \enquote{Client} according to \gls{oidc} terminology denotes an \gls{sp} according to the common \gls{sso} terminology -- a service, which can be accessed by the End-User.
In order to be compliant to the terminology within the OAuth and \gls{oidc} specification, we will use the term \enquote{Client} from now on.

\textbf{The \glsfirst{op}} acts as a \gls{ttp}/\gls{idp} towards Client and End-User, handles End-User authentication and issues an authentication token containing a specific set of claims proving the identity of the End-User.
Additionally, an authorization token can be issued, in order to authorize the Client to access End-User's resources.

\subsection{\gls{op} Endpoints}
\label{sec:basics:endpoints}

Within the \gls{oidc} Core specification~\cite{OpenIDFoundation2014} the following endpoints on the \gls{op} are defined and their relation to the according \gls{sso} phases is depicted in \autoref{fig:OIDC_Phases}:

\begin{figure*}[!ht]
	\centering
	\includegraphics[width=0.90\textwidth]{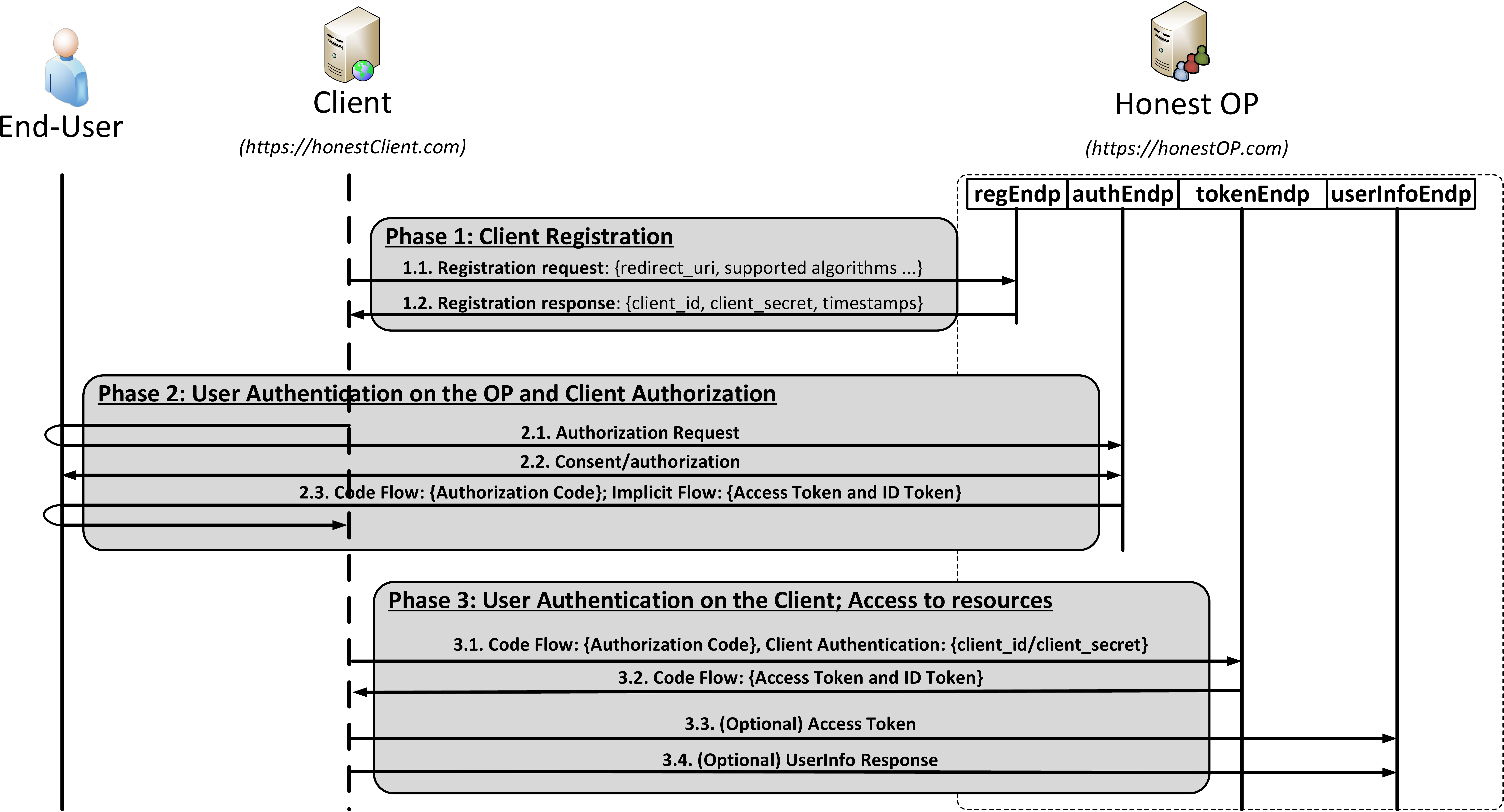}
	\caption{Information flow in \gls{oidc} containing three phases: Client Registration, User Authentication on the OP and Client Authorization, and User Authentication on the Client.}
	\label{fig:OIDC_Phases}
\end{figure*}

\begin{enumerate}
 \item \emph{Registration Endpoint (regEndp):}
 In order to use \gls{oidc} services for authentication, a Client has to register on the \gls{op}.
 For this registration, the Client accesses this URL \emph{regEndp}, e.g., \texttt{https://google.com/register}.

 \item \emph{Authorization Endpoint (authEndp):}
 In order to execute the Authentication Request of the Client, the End-User has to be redirected to the \emph{authEndp} of the \gls{op}, e.g., \texttt{https://login.google.com/}.
 Here, the End-User has to authenticate to the \gls{op} via a corresponding authentication process and authorize the Client to access the requested resources.

 \item \emph{Token Endpoint (tokenEndp):}
 The Client communicates with the \emph{tokenEndp}, e.g., \texttt{https://google.com/consume-token}, in order to obtain the \gls{id_token} described in \autoref{sec:idToken} and authenticate the End-User. In addition, an \texttt{access\_token} can be sent to the Client in order to authorize the access to restricted resources.
 This communication is done directly between Client and \gls{op} (without involving the End-User).

 \item \emph{UserInfo Endpoint (userinfoEndp):} returns information about the authenticated End-User like email, address, phone, gender etc. In order to access the resources the Client uses an Access Token obtained through the \gls{oidc} authentication.
\end{enumerate}

\subsection{Information Flow}

\gls{oidc} contains three phases, as shown in \autoref{fig:OIDC_Phases}.
In this section we explain the information flow during the different phases.

\paragraph{Phase 1: Client Registration}
The Client initially communicates with the \gls{op}'s registration endpoint (\emph{regEndp}).
It submits the \emph{domain(s)}, where the Client is deployed, for example \url{https://clientA.com}.
The \gls{op} then generates a random \gls{client_id}/ \gls{client_secret} pair, stores them both together with the domain as a triplet, and sends the credentials to the Client. %
The Client stores the same information in order to use it during phases 2 and 3.

The Client's registration is mostly processed only once and is usually done via the web interface of the \gls{op}, for example by the domain administrator or the developer of the Client.
Thus, the registration needs user interaction, causes management overhead, and cannot be executed automatically.

\paragraph{Phase 2: User Authentication on the \gls{op}}
In the context of delegated authentication the Client redirects unauthenticated End-Users to the Authorization service endpoint on the \gls{op} (\emph{authEndp}). 
Subsequently, the End-User authenticates to the \gls{op} using his credentials.
Then, the \gls{op} generates an authorization \gls{code} and sends it to the End-User. The \gls{code} is an intermediary between the End-User and the Client. By sending the \gls{code} to the Client, the End-User authorizes the Client to access restricted resources.

Once received, the Client uses the \gls{code} to retrieve the authentication token (\gls{id_token}), containing user's identity, and optionally the \gls{access_token} granting access to restricted resources.

\paragraph{Phase 3: User Authentication on the \gls{op} -- ID and Access Token}
Once the Client receives the \gls{code}, it sends it to the Token Service endpoint on the \gls{op} (\emph{tokenEndp}). 
In the same message, the Client sends its credentials (\gls{client_id} and \gls{client_secret}, cf.\ Phase 1) and authenticates to the \gls{op}.

The \gls{op} responds with the \gls{id_token} and possibly the (optional) \gls{access_token}.
Once the \gls{id_token} is received, the Client verifies it and subsequently authenticates the End-User.

The optional \gls{access_token} is part of the OAuth protocol flow and it authorizes the Client to  access restricted resources of the End-User on the \gls{op}.

\subsection{ID Token}
\label{sec:idToken}

The \gls{oidcp} is basically an extension of \gls{oauth} by adding an \gls{id_token}.
The \gls{id_token} is a security token containing claims about the identity of an End-User by an \gls{op}, proving the End-User's identity to a Client.
Its data structure is represented as a \gls{jwt}~\cite{Jones2014}.
In order to provide authenticity as well as integrity of the token, the \gls{op} is responsible for signing it using \gls{jws}~\cite{Jones2014a}.

\begin{lstlisting}[style=json,caption={An example of ID Token as JSON object.},label={lst:id_token}]
Header:   { "alg": "HS256" }
Body: 	  {
	  "iss": "http://openidConnectProvider.com/",
	  "sub": "user1",
	  "exp": 1444148908,
	  "iat": 1444148308,
	  "nonce": "40c6b33b9a2e",
	  "aud": "http://client.com/",
	  }
Signature: AF45JF93LKD76D....
\end{lstlisting}

A signed \gls{id_token} consist of three parts: Header, containing information regarding the used cryptographic algorithms, Body, including information needed for the authentication of an End-User, and Signature providing the authenticity and integrity of the \gls{id_token}.
 
\paragraph{Identity of the End-User}
The identity of the user consists of two parts: (1.) \emph{issuer} and (2.) \emph{subject}.
The \gls{iss} (cf. \autoref{lst:id_token}: \texttt{iss}, Line 3) is a mandatory identity claim identifying the originator (the \acrlong{op}) of the \gls{id_token}, for example \url{https://www.myOpenIDProvider.com}.
The \gls{sub} (\texttt{sub}, Line 4) is a mandatory identity claim that specifies the End-User's identifier and is consumed by the Client. 
Issued by the \acrlong{op} (\gls{iss}), it has to be locally unique and never reassigned (e.g., \texttt{alice@myOpenIDProvider.com}).
It is essential to note that \emph{both} values -- \gls{iss} and \gls{sub} -- must be used to uniquely identify the End-User.

\paragraph{Timestamps and Freshness}
The claims \gls{iat} (\texttt{iat}, Line 6) and \gls{exp} (\texttt{exp}, Line 5) define the creation and expiration times of the token.
The \gls{nonce} (Line 7) claim is a randomly chosen String value, sent by the Client within the Authentication Request and passed through unmodified to the \gls{id_token}, used to mitigate replay attacks.

\paragraph{Audience Restrictions}
The \gls{aud} (\texttt{aud}, Line 8) is a mandatory claim specifying the audience(s) that this \gls{id_token} is intended to be used for (e.g.\ \texttt{https://clientA.com}).
It must, at least, contain the \gls{client_id} of the Client which requests the token.

%% file: sections/attackerModel.tex
\section{Security Model}
\label{sec:models}

This section will give a detailed description of the security model used in the analysis of the \gls{oidcp}.

\subsection{Assumptions}

We make the following assumptions for the analyzed systems:

\begin{itemize}
\item \emph{Secure \acrshort{tls} channels}:
A huge proportion of the security of \gls{oidc} is based on the assumption that \gls{tls} is used to secure the communication between the involved parties.
Naturally, we follow this approach and assume the corresponding \gls{tls} channels to be secure.

\item \emph{Uncompromised software}:
All software used by the End-User is assumed to be \emph{uncompromised}.
This especially holds for the \acrlong{ua} and the operating system -- we assume that no malicious web browser plugins and that no keyloggers etc., are active on the End-User's system.
We additionally assume that the Client and the \gls{op} can also not be compromised.
For example, we assume that we do not have any other access except for their publicly available website (e.g. we do not have shell access).

\item \emph{No impersonation towards the End-User}:
The attacker controls his own webservers and services, but we assume that he does not impersonate legitimate web applications.
We thus assume that the End-User can neither be tricked into accepting attacker generated \gls{tls} certificates as valid certificates for genuine Clients, nor will the End-User react to Phishing mails claiming to originate from the legitimate Client.

In short, we assume the attacker must not able to impersonate a legitimate Client towards the End-User in any meaningful way.
\end{itemize}

\subsection{Capabilities of the Attacker}

The attacks to-be-introduced in this work have been strictly verified in the \emph{web attacker} model \cite{Barth2009}.
In contrast to the network-based attacker model (also called the \emph{cryptographic} attacker model), the web attacker does not have full control over the network and thus is unable to eavesdrop on or manipulate network connections.

He is, however, able to use a \gls{ua} or a custom \gls{http} client to send arbitrary \gls{http} requests to every publicly available web application in the web (including the Client and the \gls{op}) and subsequently receive its responses.
For tests within live implementations the attacker is able to register as many accounts on a specific Client or \gls{op} as he wishes.

Furthermore, the attacker can use links (e.g., sent via email) or web-blog commentaries to lure the victim into opening a (manipulated) \gls{uri} to, for example, conduct \gls{csrf} attacks.

\subsection{Attacker Goals}
\label{sec:attackerGolas}

The scope of this paper are attacks against the \gls{enduser} and attacks against the Client.

\paragraph{Attacking the End-User}
Attacks on the End-User are focusing on token theft.
In \gls{oidc}, there are three different tokens:
The \gls{code}, the \gls{id_token} and the \gls{access_token}.
Leakage of any of them can allow an attacker to get unauthorized access to restricted resources.

\paragraph{Attacking the Client}
Attacks on the Client have different surfaces and can be categorized in two groups.

The first group contains impersonation attacks.
In Phase 1 of the \gls{oidcp} -- during the registration -- the Client receives the \gls{client_id} and the \gls{client_secret}.
Both parameters are used for the Client's authentication to the \gls{op}.
If these credentials are compromised, the attacker can use them to impersonate the Client.
In Phase 2 of the protocol, the Client receives at least one token.
An attacker can then send manipulated tokens to the Client in order to impersonate different End-Users.

The second group contains classical attacks on \glspl{wa}. These attacks include \gls{dos} techniques as well as code injection attacks, like \gls{xss} or \gls{sqlinjection}.
Please note that in our work, we consider this group of attacks only in conjunction of the \gls{oidcp}. Thus, only attacks that are directly initiated through protocol messages are investigated.

%% file: sections/evaluationGap.tex
\section{Gap in Security Evaluations}
\label{sec:novelAnanalyzingTechniques}
\begin{figure*}[t]
	\centering
	\includegraphics[width=1.0\textwidth]{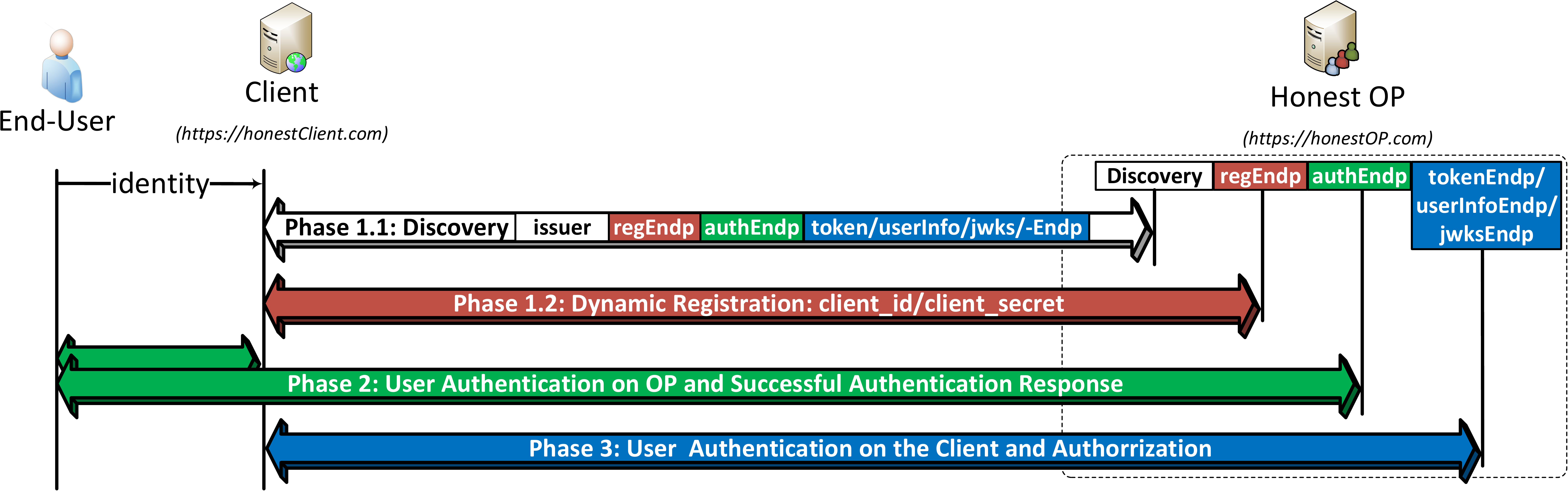}
	\caption{OpenID Connect Dynamic Registration}
	\label{fig:OIDC_DynReg}
\end{figure*}

By considering previous work regarding the security of \gls{sso} protocols, we observed that its analysis concentrates on Phase 2 and Phase 3~\cite{sun2012devil,ssoscan,demystifiedOAuthCCS14}.
Concentrating on those two phases seems plausible, because the End-User authenticates in Phase 2 and the authentication tokens are transmitted in Phase 3.
An attacker targeting Phase 2 or Phase 3 can achieve one or more of the goals defined in \autoref{sec:attackerGolas}.
As a result, previous work only revealed security vulnerabilities in Phase 2 and Phase 3. These were fixed and the specification was changed~\cite{demystifiedOAuthCCS14}.

Nevertheless, the entirety of Phase 1 has not been considered so far. 
This is reasoned by the fact that the Client Registration and Key Transport between Client and \gls{op} are usually executed manually.
For instance, the developer of a Facebook App has to visit his Facebook developer website, and click on \emph{create new App}.
Facebook will then generate a \gls{client_id} and a \gls{client_secret}.
The developer then copies them into his App configuration manually.

In contrast to this manual execution, protocols like OpenID and \gls{oidc} can also execute the Client Registration automatically.
Especially for \gls{oidc}, this issue is addressed by introducing a new approach for the Client registration:
The so called \emph{Dynamic Registration}~\cite{OpenIDFoundation2014b} allows registration to be automatic, transparent and without any user interaction.
However, an important security question raised about this development is:
\emph{How does this feature affect the security of the protocol?}

%% file: sections/secondorder.tex
\section{Second-Order Vulnerabilities in \gls{oidc}}
\label{sec:secondorder}

In this section we first describe the \gls{oidc} extensions \emph{Discovery} and \emph{Dynamic Registration} in detail. 
Then, we present security considerations regarding the usage of the both phases.
Based on the security considerations we introduce the concept of a novel class second-order vulnerabilities in \gls{sso}.

\subsection{\gls{oidc}: Discovery and Dynamic Client Registration}
The information flow during the automated Client Registration is shown in \autoref{fig:OIDC_DynReg}.
Initially, the End-User submits his \emph{identity}, for example \emph{alice@honestOP.com}, to the Client.
In order to initiate the \gls{sso} authentication, the Client needs to discover the corresponding \gls{op} controlling the identity of Alice.

\paragraph{Phase 1.1}
The Client uses the provided \emph{identity} and extracts the domain name of the \gls{op}~\cite[Section 2.1]{OpenIDFoundation2014a}.
In our example, Alice's identity is controlled by the domain \emph{honestOP.com}.
The domain name uniquely identifies the corresponding Discovery endpoint\footnote{This is usually realized by applying a modifier to the domain, e.g., \texttt{https://honestOP.com/}\emph{.well-known/webfinger}.}.

The Client sends an HTTP request to this Discovery endpoint and subsequently retrieves the \gls{op}'s configuration information including its endpoint locations:
The (Dynamic) Registration Endpoint (\emph{regEndp}), the Authorization Endpoint (\emph{authEndp}), the Token Endpoint (\emph{tokenEndp}) and further endpoints (c.f., \autoref{sec:basics:endpoints}).

\paragraph{Phase 1.2}
In Phase 1.2 (Dynamic Registration) the Client can automatically register at the \gls{op}.
For that purpose, the Client sends its own URL, for example \texttt{http://client.com}, to the \emph{regEndp} URL.
The \gls{op} responds with a \gls{client_id}/ \gls{client_secret} pair.
Finally, the Client and the \gls{op} store the credentials in their respective databases and use them during the next phases.

\subsection{Influence of the Discovery phase on the \gls{oidc} flow}

By analyzing the Discovery and Dynamic Registration phases we make the following observations:

\begin{itemize}
    \item The usage of any \glspl{op} is supported by the \gls{oidcp} without any pre-configuration, installation or manual interaction (neither on the Client nor on the  User-Agent).
    The End-User has to enter his identity on the Client, e.g. \texttt{bob@honestOP.com}, in order to start the authentication with his own \gls{op}.

    \item All discovered endpoints are URLs.
    No limitations are specified that restrict these URLs to domains, subdomains, or URL contexts.
\end{itemize}

\begin{figure}[!ht]
 \centering
 \includegraphics[width=0.95\linewidth]{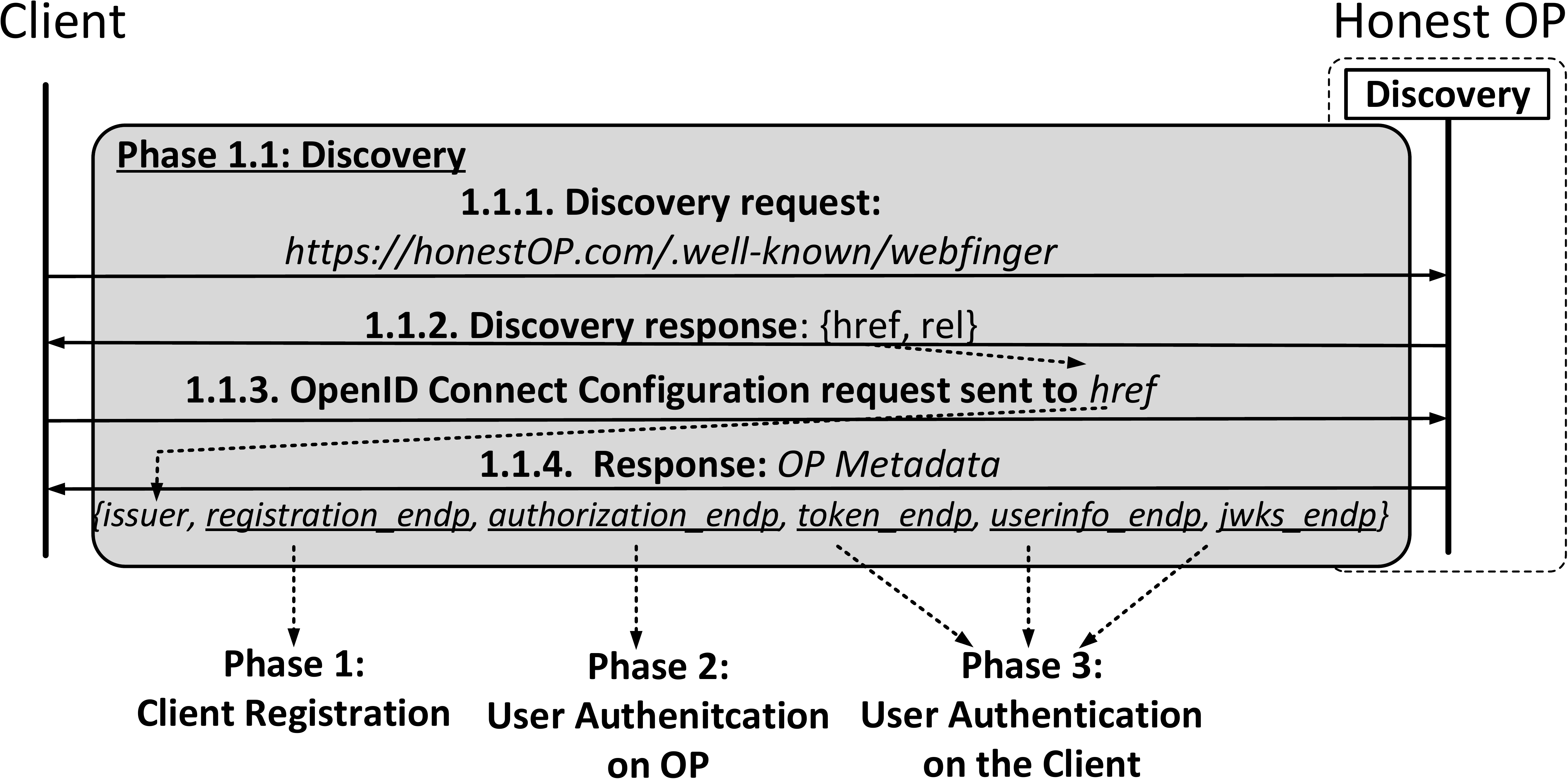}
 \caption{A detailed overview of the Discovery phase revealing how the metadata received by the Client influences the next \gls{oidc} phases.}
 \label{fig:metadataInfluence}
\end{figure}

Based on our observations, we discovered that we can trigger any Client supporting Discovery and Dynamic Registration to use our custom \gls{op} for authentication.
Thus, we control the data sent to the Client and used in the following phases of the protocol flow. In \autoref{fig:metadataInfluence} we present how the retrieved information during the Discovery phase influences the \gls{oidc} phases.

The first two messages are used to discover the URL where the metadata of the \gls{op} is stored based on the identity entered by and End-User. The \emph{Discovery request} (1.1.1) is an HTTP message sent to the \gls{op}'s discovery service (e.g.\emph{ \url{https://honestOP.com/.well-known/webfinger}}). The response is a JSON message containing two parameters: 
\begin{inparaenum}
	\item \emph{href}, which points to the metadata of the \gls{op} and 
	\item \gls{rel}, which identifies the type of service (e.g. \emph{\url{http://openid.net/specs/connect/1.0/issuer}}).
\end{inparaenum}

Consequentially, a new HTTP request is sent to the URL specified via the \emph{href} parameter. The response contains the metadata with all information regarding the \gls{op}: endpoints, supported authentication flows, supported algorithms for signing and encrypting messages, public keys of the \gls{op} etc.

\autoref{fig:metadataInfluence} depicts the relation between the endpoints received in the last messages of the Discovery phase and the \gls{oidc} phases.
The \emph{regEndp} will be used by the Client in order to register the Client on the \gls{op} and receive the client credentials (e.g. \emph{client\_id/client\_secret}).

The \emph{authEndp} points to the Authorization server responsible for the authentication of the End-User. Noteworthy is the fact that only the Authorization server is visible for the End-User during the authentication. All other endpoints are called by the Client directly and thus cannot be seen by the End-User.

The next three endpoints \emph{tokenEndp}, \emph{userInfoEndp} and \emph{jwksEndp} are used in the last Phase (Phase 3) of the protocol: the End-User authentication. 

\subsection{Security considerations}

\gls{oidc} supports the usage of custom \glspl{op}.
For that purpose, a Client uses the Discovery and the Dynamic Registration Phase to retrieve the \gls{op}'s configuration information including the endpoints \emph{regEndp}, \emph{authEndp}, \emph{tokenEndp}/\emph{userInfoEndp}/\emph{jwksEndp} and registers on it. 

Please note that a malicious Discovery service can freely choose all these parameters (cf., Phase 1.1 in \autoref{fig:OIDC_DynReg}).
By this means, the malicious Discovery service can influence 
\begin{inparaenum}
	\item on which URL the Client registers (\emph{regEndp}),
	\item which URL is used by the End-User to authenticate (\emph{authEndp}),
	\item and to which URL the token will be finally sent (\emph{tokenEndp/userInfoEndp}).
\end{inparaenum}

\subsection{Second-Order vulnerabilities in \gls{oidc}}
Based on the security considerations, we developed a new class of attacks referred to the second-order vulnerabilities. 
This new class differ from the class of conventional second-order vulnerabilities in the context of \gls{xss}, \gls{sqlinjection} and \gls{dos}.
To clarify the difference we first explain how second-order vulnerabilities are exploited in general and then we introduce second-order vulnerabilities in distributed systems like \gls{sso} protocols.

\paragraph{Web application}
Common second-order vulnerabilities in web applications as shown in~\cite{Dahse2014,Olivo2015} have only three entities involved:
\begin{inparaenum}
	\item the attacker acting with his browser,
	\item the server hosting the web application and
	\item optionally another End-User (the victim).
\end{inparaenum}
In case of \gls{dos} attacks as shown in~\cite{Olivo2015}, there is no third entity involved, since the victim is the server hosting the web application itself.

\begin{figure}[!ht]
 \centering
 \includegraphics[width=0.7\linewidth]{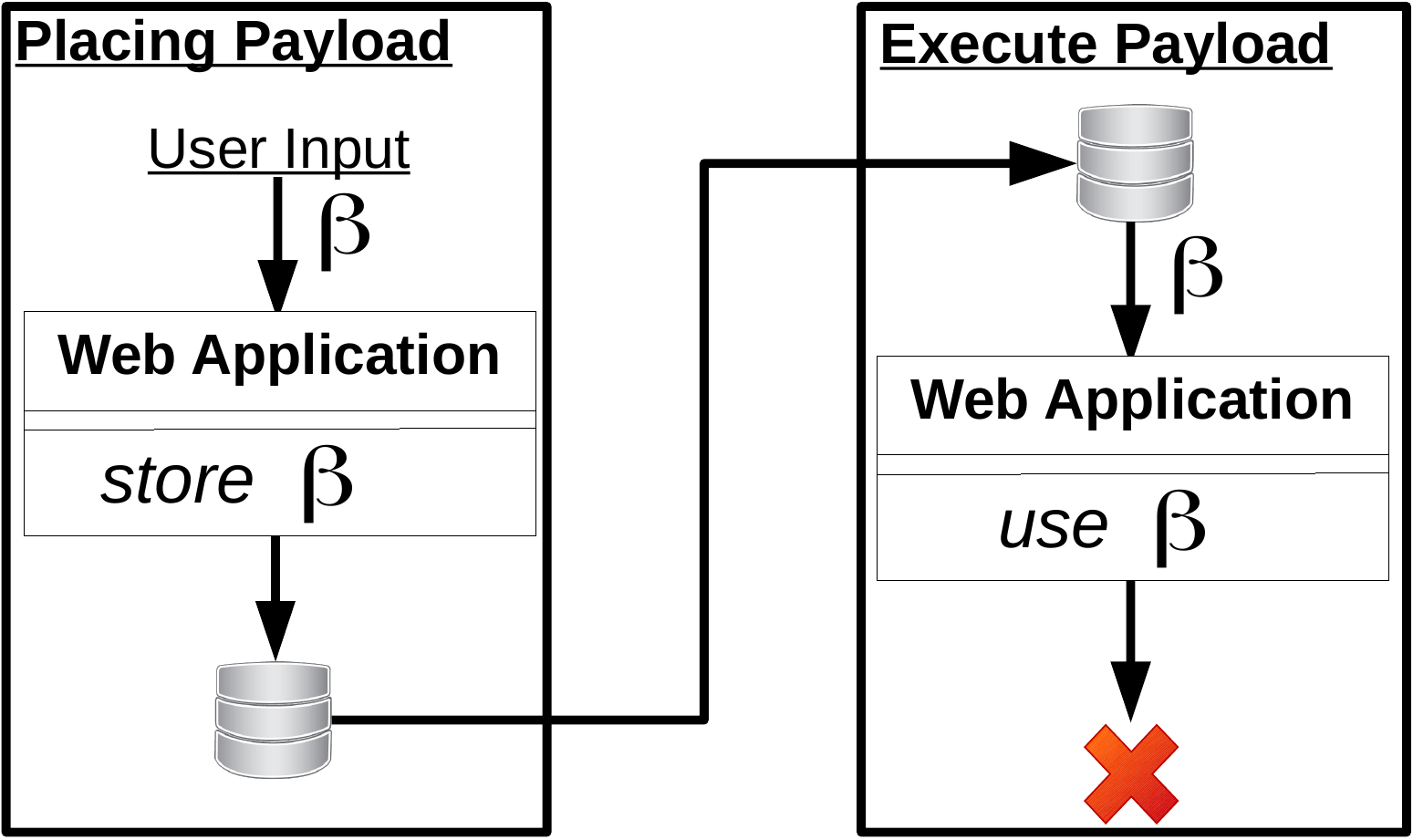}
 \caption{Data flow of a conventional second-order vulnerability on an web application. The attack vector $\beta$ is first placed, and later on executed leading to security issues.}
 \label{fig:webapp2order}
\end{figure}

\autoref{fig:webapp2order} shows an example of a second-order vulnerability on a \gls{wa}.

During the first step, the user input containing an attack vector will be stored in the database of a \gls{wa}. In this step, no attack, but its preparation will be processed.
Later on, the stored attack vector will be pulled from the database and executed leading to an \gls{sqlinjection}, \gls{xss} or even \gls{dos}.

\paragraph{\gls{oidc}}
Second-order vulnerabilities in \gls{sso} protocols are more complex than on web applications, since the attack consists of multiple steps and messages exchanged between different participants, for example, End-User and Client, Client and \gls{op}, and End-User and \gls{op}.
Due to the nature of \gls{sso}, we have more entities:
\begin{inparaenum}
	\item the End-User who wants to login. This can be either a benign End-User or the attacker;
	\item the Client which is the main target of our attacks;
	\item a honest \glsfirst{op};
	\item an attacker hosting his own service on the Internet. We will use this service to host a \gls{mds} later on.
\end{inparaenum}

\begin{figure}[!ht]
	\centering
	\includegraphics[width=0.95\linewidth]{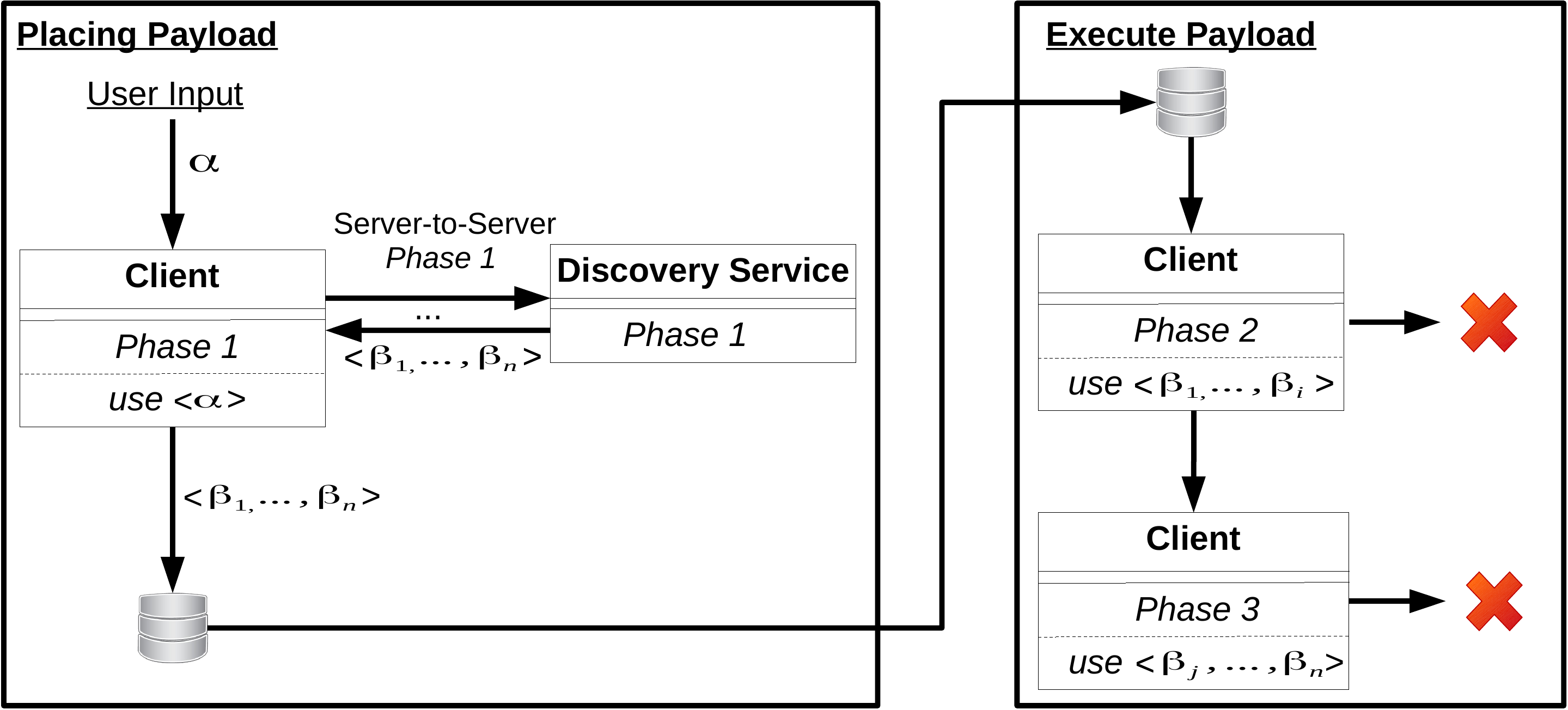}
	\caption{Data flow of a second-order vulnerability in \gls{oidc}. User's input $\alpha$ triggers the authentication. Within the Server-to-Server communication in \emph{Phase 1}, the attack vectors will be placed. These will be used later during \emph{Phase 2} or/and \emph{Phase 3}. Please note that the contacted discovery services depends on the value of $\alpha$.}
	\label{fig:sso2order}
\end{figure}

\autoref{fig:sso2order} shows the data flow within a second-order vulnerability in a \gls{sso} protocol.
Initially, the attacker stores the attack vectors. The main difference is that the user input $\alpha$ is not the attack vectors. $\alpha$ just starts the \gls{sso} authentication.
The injection of the attack vectors occurs during \emph{Phase 1} of the protocol -- the Discovery phase within a Server-to-Server communication. The attacker returns data used later on during the protocol. 
In case of \gls{oidc} this is a metadata file containing endpoints of the \gls{op}, supported protocol flows and supported cryptographic algorithms.

Later on, the stored attack vectors $\beta_{1}, ..., \beta_{n} $ will be loaded.
Please note that each attack vector can be used during different \gls{sso} phases. Thus, $\beta_{1}, ..., \beta_{i}$ are used during Phase 2 of the protocol resulting. 
They can either lead to successful completion of this phase or to an successful attack.
The further attack vectors $\beta_{j}, ..., \beta_{n}$ will be used in Phase 3 and lead to security issues.

To summarize, in Step 1 the attacker can place a harmless\footnote{\emph{Harmless in this context means that the payload is not directly executed.}} payload on the Client in such a way that the further communication process between the participants is influenced resulting in the following issues:
\begin{enumerate}
 \item The attacker gets access to resources owned by an benign End-User (\autoref{sec:brokenenduserauth}).
 \item The attacker gather sensitive information regarding the Client (\autoref{sec:ssrfOIDC})and thus breaking privacy?.
 \item The attacker injects further attack vectors like \gls{xss} or \gls{sqlinjection} (\autoref{sec:injectionattacks}).
 \item The further communication process between Client and End-User or Client and \gls{op} is slowed down significantly due to a \gls{dos} attack (\autoref{sec:dosattacks}).
\end{enumerate}

%% file: sections/attacks.tex
\section{Malicious Endpoints Attacks}
\label{sec:maliciousendpointattack}

This section describes four different attacks, which belong to the class of \gls{mep} attacks.
All attacks use the \emph{\gls{mds}} and influence the \gls{oidc} flow.
Since each attack pursues different goals, we describe for each attack the main goal, the setup including the attacker model and the attack itself.

\begin{figure*}[t]
 \centering
 \includegraphics[width=1.0\textwidth]{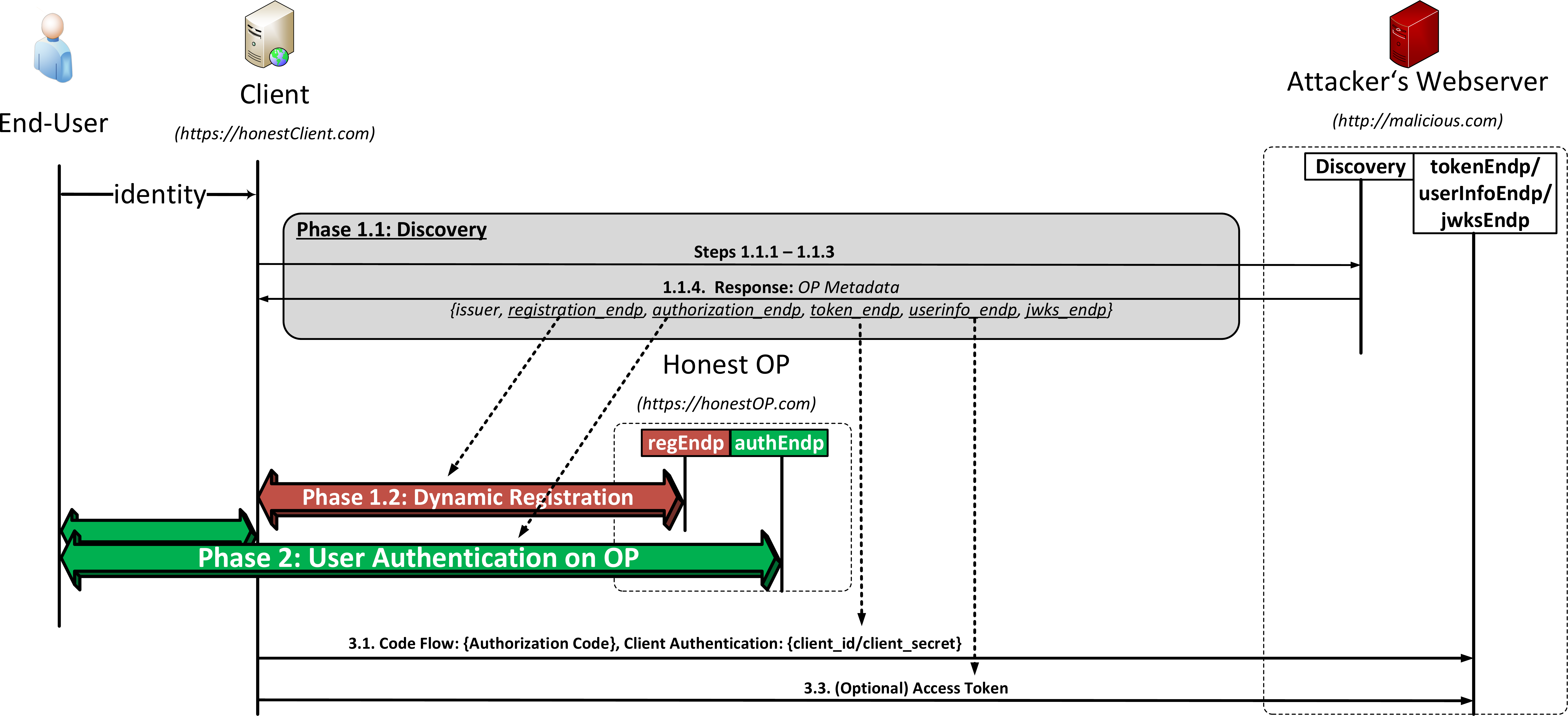}
 \caption{Malicious Endpoints attack: Attacker's Discovery service sets the endpoint variables in a specific way, such that the secret tokens sent in the third phase are seamlessly distributed to the attacker's server.}
 \label{fig:maliciousEndpoints}
\end{figure*}

\subsection{Broken End-User Authentication}
\label{sec:brokenenduserauth}
The idea behind the attack is to influence the information flow in the Discovery and Dynamic Registration Phase in such a way that the attacker gains access to sensitive information.
The attacker pursues the theft of the credentials between the honest \gls{op} and the honest Client.
Additionally, he steals a valid \gls{code} authorizing the Client to access End-User's resources on the honest \gls{op}.

\paragraph{Setup}
The basic setup for the attack is as follows:

\begin{itemize}
\item The End-User (\emph{victim}) has an active account on the genuine \emph{honest Client}.
We assume that the End-User trusts this Client and the Client follows the \gls{oidcp} rules.

\item The End-User is registered at the \emph{honest \gls{op}} on the domain \emph{\url{https://honestOP.com}}.
The End-User trusts this \gls{op} and the \gls{op} also follows the \gls{oidcp} rules.

\item To perform the attack, the attacker has to set up his own Discovery service running on the domain \emph{\url{http://malicious.com}}.
This Discovery Service acts maliciously in that it deviates from the \gls{oidcp} flow as described in \autoref{fig:OIDC_DynReg}.
Note that there is no need to disguise \emph{\url{http://malicious.com}} as the regular Discovery service belonging to the honest \gls{op} in any way.

\item According to the attacker model, the attacker does not hold any control over the honest Client, the End-User, the honest \gls{op} or the network traffic between these instances.
The attacker is able to send an HTTP request through End-User's browser, e.g. by embedding an image in a benign HTML website that causes the browser to automatically issue a request when the website is viewed.

\end{itemize}

\paragraph{Attack description}
In the following, we describe the attack protocol flow, which we depicted in \autoref{fig:maliciousEndpoints}.

\emph{Phase 1.1 - Injecting malicious endpoints}
The attacker's intention in the first phase is to force a valid Client to use the attacker's \gls{mds}.
For this purpose, he constructs a malicious link and stores it on a benign website, e.g. in a web forum.
For example, this can be a link to the valid Client containing an identity \emph{alice@malicious.com}.

\begin{lstlisting}[frame=shadowbox,xrightmargin=1.1em,caption={Endpoints returned by the \gls{mds}},label=lst:maliciousEndpoints,style=base]
@issuer:    	http://malicious.com@
regEndp:   	https://honestOP.com/register
authEndp:  	https://login.honestOP.com/
@tokenEndp: 	http://malicious.com@
@userInfoEndp:	http://malicious.com@
\end{lstlisting}

By visiting the website containing the malicious link, an HTTP request will be sent to the Client through the End-User's (victim's) browser.
Consequentially, the Client starts a discovery phase with the \gls{mds} \url{http://malicious.com}.
The Client sends a request to determine the corresponding endpoints.
The attacker's Discovery service responds with the following values, initiating the actual attack:

\emph{Phase 1.2 -- Dynamic Registration}
In the next step, the Client accesses \emph{regEndp} for the Dynamic Registration.
It sends a registration request to \url{https://honestOP.com/register} and receives a \emph{client\_id} and \emph{client\_secret} in the response.

\emph{Note:} The Client automatically starts the Dynamic Registration, even if it is already registered on the honest \gls{op}.
The reason for this behavior is that the Client believes that \url{http://malicious.com} is the responsible \gls{op}, since it is not known from previous authentication procedures.
Thus, \url{http://malicious.com} is a new \gls{op} for the Client and it starts the registration procedure.

\emph{Phase 2 -- End-User Authentication and Authorization}
In the next phase, the Client redirects the End-User to \emph{authEndp}, \url{https://login.honestOP.com/}, where the End-User has to authenticate himself and authorize the Client.
The End-User is not able to detect any abnormalities in the protocol flow:
Phase 1.1 and Phase 1.2 cannot be observed by the End-User, and in Phase 2 the End-User will be prompted to authenticate to the honest \gls{op} and authorize the honest Client, both of which he knows and trusts.
Thus, the End-User authorizes the Client and the \gls{op} generates the \gls{code}, which is sent to the Client.

\emph{Note:} Phase 2 exactly follows the original \gls{oidcp} flow -- there are no parameter manipulations, no redirects to malicious websites and no observation of the network traffic between the End-User, the honest \gls{op} and the Client.
Thus, the attack started at the beginning of the protocol flow can be neither detected nor prevented by any of the participants at this point.

\emph{Phase 3 -- The Theft}
In dependence of the protocol flow, Code or Implicit, the messages sent to the attacker differ.

Within the \emph{Code flow} the Client redeems the received \gls{code} from the previous phase: It sends the \gls{code} together with the corresponding Client's credentials received during the Dynamic Registration (\emph{client\_id/ client\_secret}) to the \emph{tokenEndp} originally specified by the \gls{mds} -- in this example \url{http://malicious.com}, see \autoref{lst:maliciousEndpoints}.

Since the \emph{Implicit flow} does not use the \emph{tokenEndp}, the attacker is not able to receive the information send in phase 2. However, he can use another malicious endpoint -- \emph{userInfoEndp} used in Step 3.3 in \autoref{fig:maliciousEndpoints} to retrieve further information about the authenticated user. In the request, the Client sends a freshly generated Access Token. As a result, the attacker receives this Access Token and is able to access the authorized resources on the \gls{op}.

\subsection{\glsfirst{ssrf}}
\label{sec:ssrfOIDC}
A \gls{ssrf} attack describes the ability of an attacker to create requests from a vulnerable web application to the application's Intranet and the Internet. Usually, \gls{ssrf} is used to attack internal services placed behind a firewall and not accessible from Internet.
In context of \gls{oidc}, the \gls{mds} can be used to start such attacks in order to 
\begin{inparaenum}
    \item gather information about the Intranet infrastructure of the Client, and 
    \item disseminate attack vectors.
\end{inparaenum}

\paragraph{Setup}
The attacker sets up a \gls{mds} returning endpoints called by the Client during the protocol flow. The endpoints are URL strings specifying protocol (http(s), ftp, smb etc.), port, path, and parameters.
Since there are no restrictions regarding the URLs, these can point to the Intranet infrastructure of the Client.
The Client will use these URLs and performs HTTP GET requests on them.
In this manner, the Client can, for example, be enforced to invoke internal REST-based web services.
This capability of the attacker is considered by the attacker model, since the attacker is able to use his \gls{ua} and send arbitrary HTTP requests to every publicly available domain. 
Thus, he can cause the Client to establish connection with the \gls{mds}.

\paragraph{Attack description}
In comparison to the Malicious Endpoint attack, now the attacker initiates the \gls{oidc} authentication on the Client by entering his identity (e.g. \emph{oskar@malicious.com}). Thus, no \gls{csrf} attack is needed.
In the end of the Discovery phase, the \gls{mds} returns the malicious endpoints called during the different phases of the protocol.
Previous researches reveal how the execution of URLs can be used to 
\begin{inparaenum}
	\item connect and execute commands on different services like Memcached, 
	\item Port scanning and 
	\item data retrieving~\cite{ssrf_blackhat2012,ssrfbible}.
\end{inparaenum}

\subsection{Code Injection Attacks}
\label{sec:injectionattacks}
User's input sent through the web interface of the Client is usually treated as untrusted and thus filtered to prevent attacks like Cross-Site-Scripting (\gls{xss}) and \gls{sqlinjection}.
In order to bypass the existing filter an attacker can use other channels to inject the attacks vectors -- for instance within the server-to-server communication in Phase 3.

\paragraph{Setup}
The attacker configures his server to inject malicious content in the messages returned in Step 3.2 (e.g. in the ID Token) or in Step 3.4 (informations about the authenticated user), which are sent to Client in Phase 3 (see \autoref{fig:OIDC_Phases}). Please note that the ID Token and Access Token returned by the malicious server are valid according to the specification, since there are no restrictions regarding the values of parameters like ``sub'', ``name'' or ``preferred\_username''.

\paragraph{Attack description}
Initially, the attacker starts the \gls{oidc} authentication on the Client by entering his identity (e.g. \emph{oskar@malicious.com}). He proceeds with the protocol execution until Steps 3.1 and 3.3.
The malicious server then responds with valid tokens (ID Token and Access Token) containing the attack vectors.
A toy example of such attack vector is shown in \autoref{lst:xssAccessToken} where an \gls{xss} attack vector is injected into the field presenting the name of an authenticated user in Step 3.4.

\begin{lstlisting}[style=json,caption={An example of an \gls{xss} attack vector hidden in the "name" filed within an Access Token.},label={lst:xssAccessToken}]
{
  "sub":"90342.ASDFJWFA",
  "name":"<script>alert(1)</script>",
  "preferred_username":"admin",
  "email":"bob@malicious.com",
  "email_verified":true
}
\end{lstlisting}

Now, the placed \gls{xss} attack vector is stored in the web application (\emph{persistent \gls{xss}}).
Other webpages on the client will use it, for example on a guestbook page, and embed the code, so that other page visitors get harmed.
The same schema can be used to place \gls{sqlinjection} attack vectors.

\subsection{\glsfirst{dos} Attacks}
\label{sec:dosattacks}
By applying \gls{dos} attacks the attacker allocates resources on a Client and negatively affects its workflow.  
Such resources are CPU usage, network traffic or memory.
The attack can target one or multiple of these resources during the execution of \gls{dos} attack.

\begin{figure}[!ht]
	\centering
	\begin{subfigure}[b]{0.5\textwidth}
		\centering
		\includegraphics[width=0.9\textwidth]{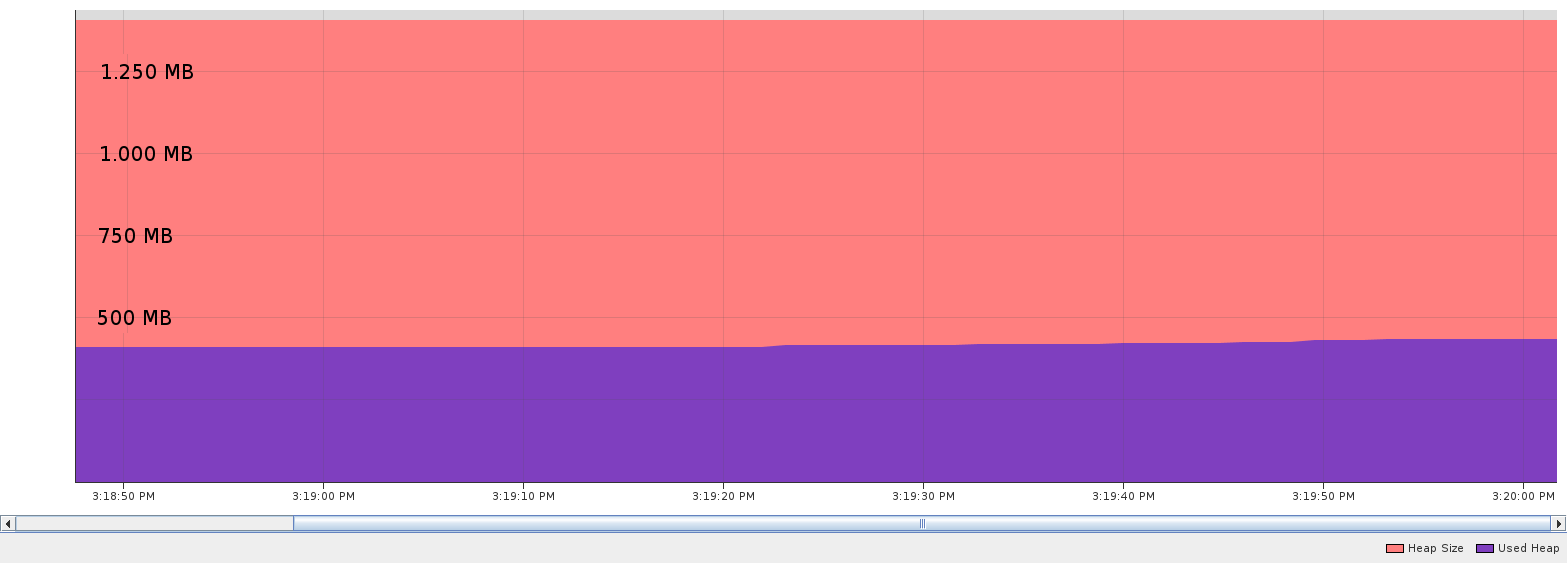}
		\caption{Memory usage on the Client within 5 parallel \gls{oidc} authentication flows to an honest \gls{op}.}
		\label{fig:normalFlow}
	\end{subfigure}
	\begin{subfigure}[b]{0.5\textwidth}
		\centering
		\includegraphics[width=0.9\textwidth]{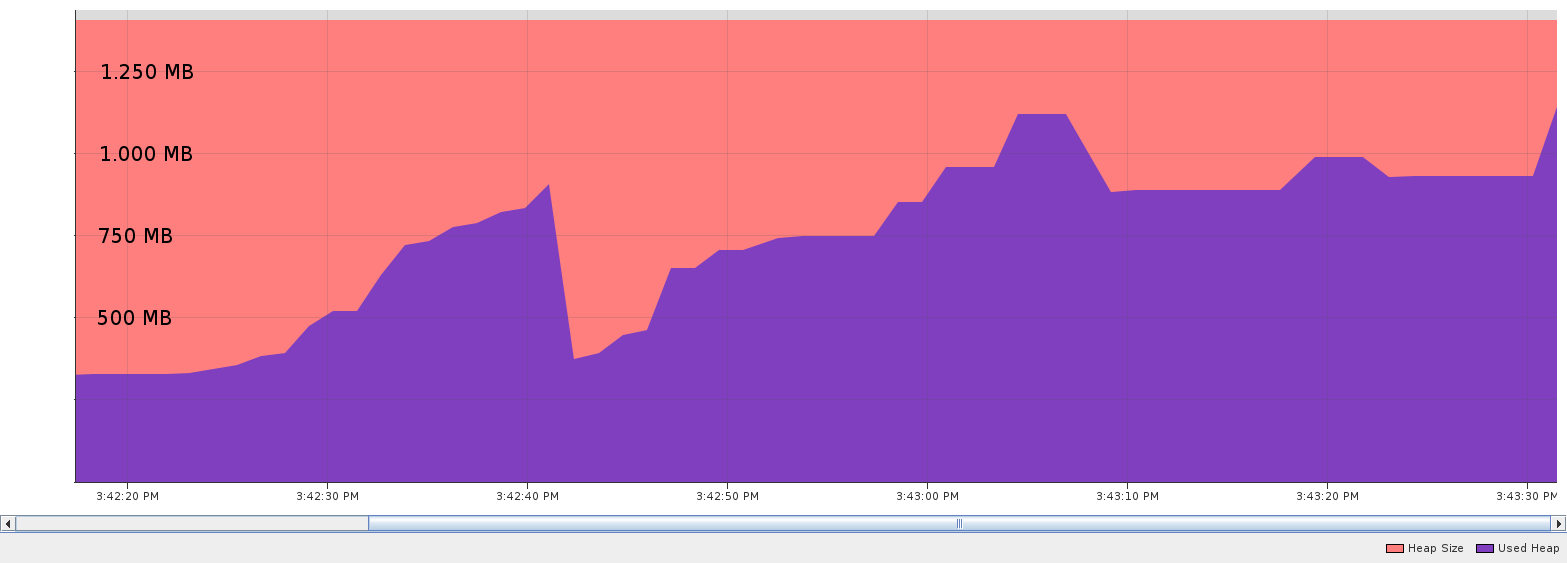}
		\caption{Memory usage on the Client within 5 parallel \gls{oidc} authentication flows to a \gls{mds} pointing to a large file (in this case, we used a Debian Linux image file with 3.7GB).}
		\label{fig:dosFlow}
	\end{subfigure}
	
	\caption{Direct comparison between the memory usage on the Client using (a) an honest \gls{op} and a (b) \gls{mds}.}
	\label{fig:dosmeasurement}
\end{figure}

\paragraph{Setup}
The setup is similar to the SSRF attack -- the attacker sets up a \gls{mds} returning endpoints called by the Client during the protocol flow.
The attacker is able to use his \gls{ua} and send HTTP request to the Client causing the Client to establish connection with the \gls{mds}.

\paragraph{Attack description}
An attack can be started by using a malicious endpoint pointing to a large data file, which will be downloaded.
The Client calls later on the malicious endpoint URL, allocates network resources as well as large amount of the memory, which will be unnecessarily used.

We provide a measurement shown in \autoref{fig:dosmeasurement} on an Apache Tomcat server with 1280 MB memory and 4x2.4 Ghz CPU.
In \autoref{fig:normalFlow} we first measured the memory usage on the Client within five parallel \gls{oidc} protocol runs with an honest \gls{op}. Once can say that almost imperceptible changes in the memory consumption occur.
In \autoref{fig:dosFlow} we repeated the same tests, but this time we used our \gls{mds} pointing to a large file. After few seconds, the memory usage increased almost threefold.  After 60 seconds, the Client was not accessible for any incoming requests.

%% file: sections/implementation.tex
\section{Implementation}
\label{sec:demo}

We implemented a web service that is publicly available on \emph{\url{http://ssoattacks.org/OIDC_MaliciousDiscoveryService/}} and it can be used by Clients for verifying the security against the attacks described in \autoref{sec:maliciousendpointattack}. In order to avoid misuse we do not provide tests for \gls{dos}, \gls{ssrf} and injection attacks.

The service contains the following informations:
\begin{inparaenum}
 \item Attack description presenting the main concept of Malicious Endpoints.
 \item A Demo showing a normal \gls{oidc} flow and additionally the broken End-User authentication attack described in \autoref{sec:brokenenduserauth}. For that purpose we configured an honest Client and an honest \gls{op}. The usage of own Clients is supported by the service.
 \item The configuration of the \gls{mds}, enabling the view on the used malicious endpoints.
 \item Database viewing all collected credentials.
\end{inparaenum}

\paragraph{Testing Clients}
Security auditors can test their Clients by using our service. 
No pre-configuration or installation of any software is needed.

\begin{figure}[!ht]
	\centering
	\includegraphics[width=0.75\linewidth]{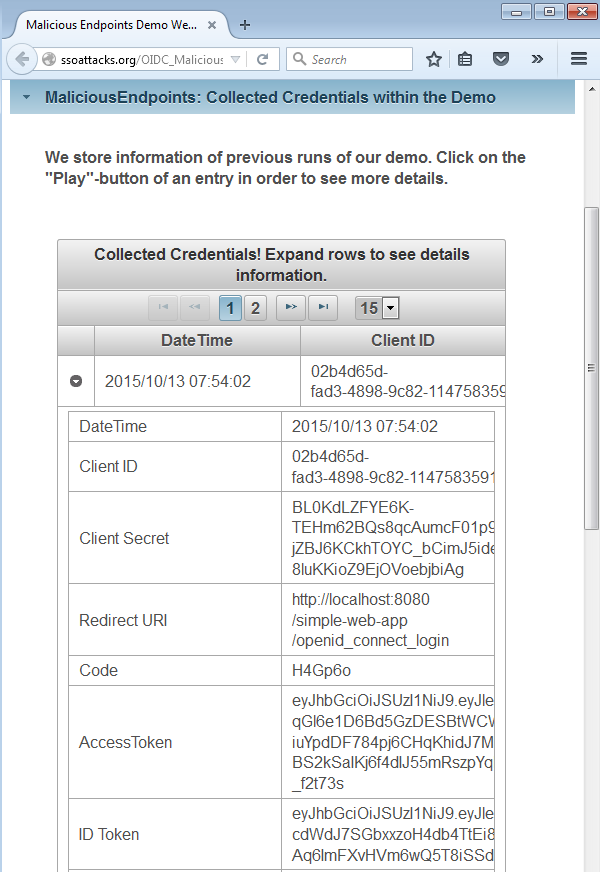}
	\caption{The implementation collects all stolen tokens and credentials and prints out a report containing the results. Thus, security auditors can test their Clients and evaluate the results of the tests.}
	\label{fig:demo}
\end{figure}

In order to start the security evaluation, the auditor has to enter the URL of our \gls{mds} on the target Client -- \emph{\url{http://ssoattacks.org/OIDC_MaliciousDiscoveryService/}}.
The Client calls the \gls{mds} and caches the metadata returned by it. The metadata contains now the malicious endpoints.
The Client proceeds with the \gls{oidc} protocol flow and starts the next phases.
At the end, the service prints out a report that includes all stolen information, see \autoref{fig:demo}, for example, the stolen \emph{code}, \emph{access\_token}, \emph{id\_token} and Client credentials.
Security auditors can use it to test arbitrary Clients and evaluate the impact of the attack.

We tested the Client applications of MITREidConnect and phpOIDC and attacked them successfully with our attacks.
An online demo Client is also available for testing, see \emph{run demo} on \url{http://ssoattacks.org/OIDC_MaliciousDiscoveryService/}.

%% file: sections/countermeasures.tex
\section{Countermeasures}

During our search for applicable countermeasures, we tried to find a solution requiring minimal changes to the protocol and to the existing implementations.
In the following, we present four possible countermeasures to our attacks and discuss the advantages and disadvantages.
Noteworthy is the fact that each countermeasure mitigates only partially the existing issues. Thus, a combination of the proposed countermeasures is needed to improve the security on the Client.

\begin{figure}[!ht]
	\centering
	\includegraphics[width=0.48\textwidth,height=100px]{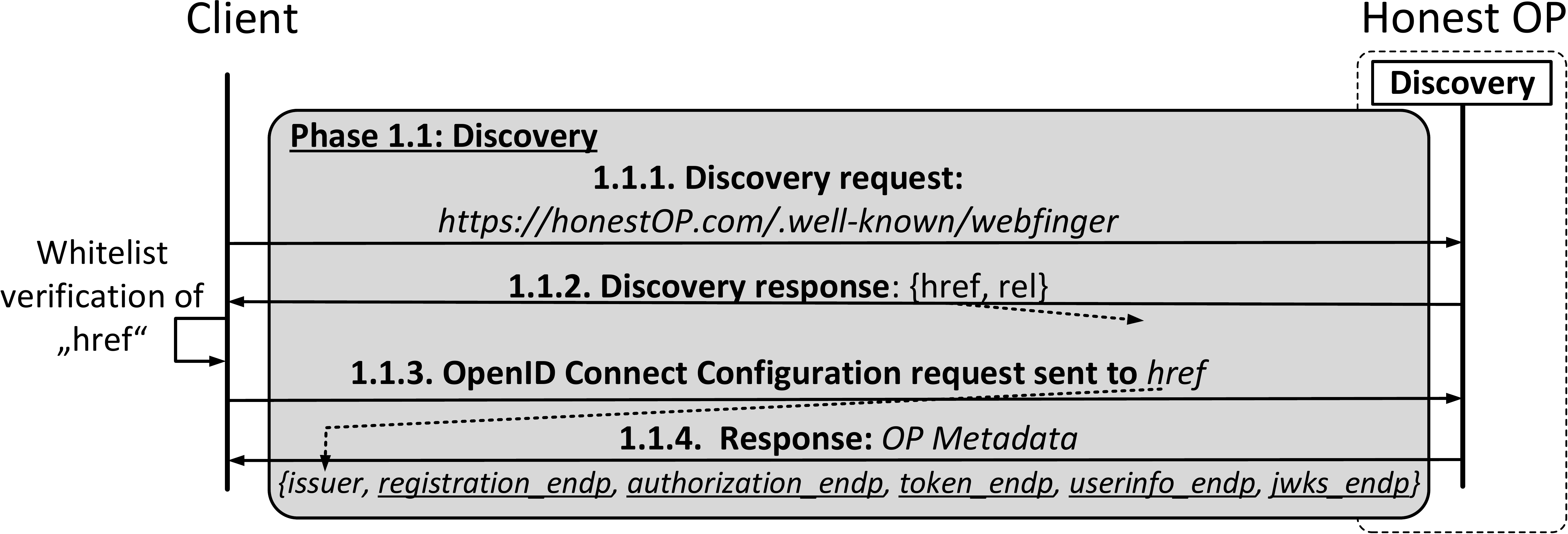}
	\caption{A whitelist verification after Step 1.1.2. during the Discovery phase. The returned \emph{href} value is compared with the entries in the database and in case that it is whitelisted the Client proceeds with the Discovery.}
	\label{fig:countermeasurewhitelist}
\end{figure}

\paragraph{\glspl{op} Whitelisting}
A suitable option to prevent the \gls{mep} attack is to whitelist the allowed \glspl{op} on Client side.
By this means, an attacker will not be able to start the authentication process with his Discovery service.
As shown in \autoref{fig:countermeasurewhitelist}, the administrator of the Client should manually whitelist the URL of each \gls{op} when it receives a discovered URL in Step 1.1.2 (the \emph{href} parameter). %

If the whitelisting approach is applied suitable, the attacker can only influence the first two messages of the discovery phase.
After Step 1.1.2, the Client compares the returned \emph{href} value with the stored values and breaks the execution in case that the URL is unknown.

Whitelisting mitigates both the \gls{mep} and the SSRF attack. This countermeasure however limits the flexibility of the Client and reduces the support of custom \glspl{op}, which, depending on the according Client, could cause problems. Additionally, the management overhead regarding the provided whitelist can lead to further problems.

\paragraph{Endpoint Restrictions}
A similar approach to prevent the attack would be to restrict the possible contents of the \emph{tokenEndp} according to the contents of the \emph{authEndp}.
For example, it could be required that the \emph{tokenEndp} MUST be located on the same domain as the \emph{authEndp} and may only differ in subdomain and/or path.
This way, an honest Client receiving a Discovery response can detect this attack and abort the protocol.

Even though this countermeasure restricts the introduced attack, it does not mitigate it completely.
In case the attacker runs his malicious Discovery service on the same Infrastructure-as-a-Service cloud environment as the honest \gls{op}, he could bypass this proposed countermeasure by using the same domain or subdomain.

\paragraph{\gls{dos} protection}
In order to prevent \gls{dos} the Client could simply do an HTTP \emph{HEAD} request --- before doing a \emph{GET} request --- and check the \emph{Content-Length} HTTP header.
In this way, the Client can retrieve the size of the file.
Please note that this will only work on benign HTTP servers.
The attacker could prepare his own webserver that responds with a wrong \emph{Content-Length} header if a \emph{HEAD} request receives.
To protect against this, the implementation should stop downloading files after receiving a specified number of Bytes (e.g. 5MB).

\paragraph{CSRF/Clickjacking protection}
Our attack to break the End-User authentication (cf. \autoref{sec:brokenenduserauth}) requires the injection of the malicious identity (e.g. \emph{alice@malicious.com}) in the first step, see \autoref{fig:maliciousEndpoints}.

To prevent this kind of attack, we propose each client to implement a proper CSRF\footnote{\url{https://www.owasp.org/index.php/Cross-Site_Request_Forgery_(CSRF)_Prevention_Cheat_Sheet}} and Clickjacking protection\footnote{\url{https://www.owasp.org/index.php/Clickjacking_Defense_Cheat_Sheet}}.

Please note that this will only prevent the broken End-User attack.
\gls{ssrf}, code injection, and \gls{dos} is still possible, because for this kind of attacks, the attacker itself sends a login request with \emph{alice@malicious.com}.

\paragraph{Client authentication via \emph{client\_secret\_jwt} or \emph{private\_key\_jwt}}
During the attack described in \autoref{sec:brokenenduserauth}, the attacker steals the \emph{client\_id/client\_secret} of the Client in step 3.1 (see \autoref{fig:maliciousEndpoints}). 
In order to mitigate the attack, the Client can use alternatively the \emph{client\_secret\_jwt} or \emph{private\_key\_jwt} flow for authentication in Phase 3~\cite[Section 9]{OpenIDFoundation2014}. 
Within both flows, the Client does not send the \emph{client\_secret} through the channel.
The client signs a JSON message by using either the \emph{client\_secret} or an asymmetric private key.  The \gls{op} verifies the message with the corresponding key and authenticates the Client.

Just signing the message does not mitigate the attack since the attacker can reply it on the honest \emph{tokenEndp}.
According to the specification the JSON message includes an \emph{audience} parameter specifying the URL of the \emph{tokenEndp}. 
During the attack the \emph{audience} points to \emph{http://malicious.com}.
Thus, the attacker can steal the \emph{code} and the signed JSON message, but he cannot replay them on the honest \emph{tokenEndp} since it will detect the different values.

Please note that \gls{dos} and \gls{ssrf} attacks are still possible and should be considered by the implementation of the Client.

\begin{figure}[!ht]
	\centering
	\includegraphics[width=0.48\textwidth,height=120px]{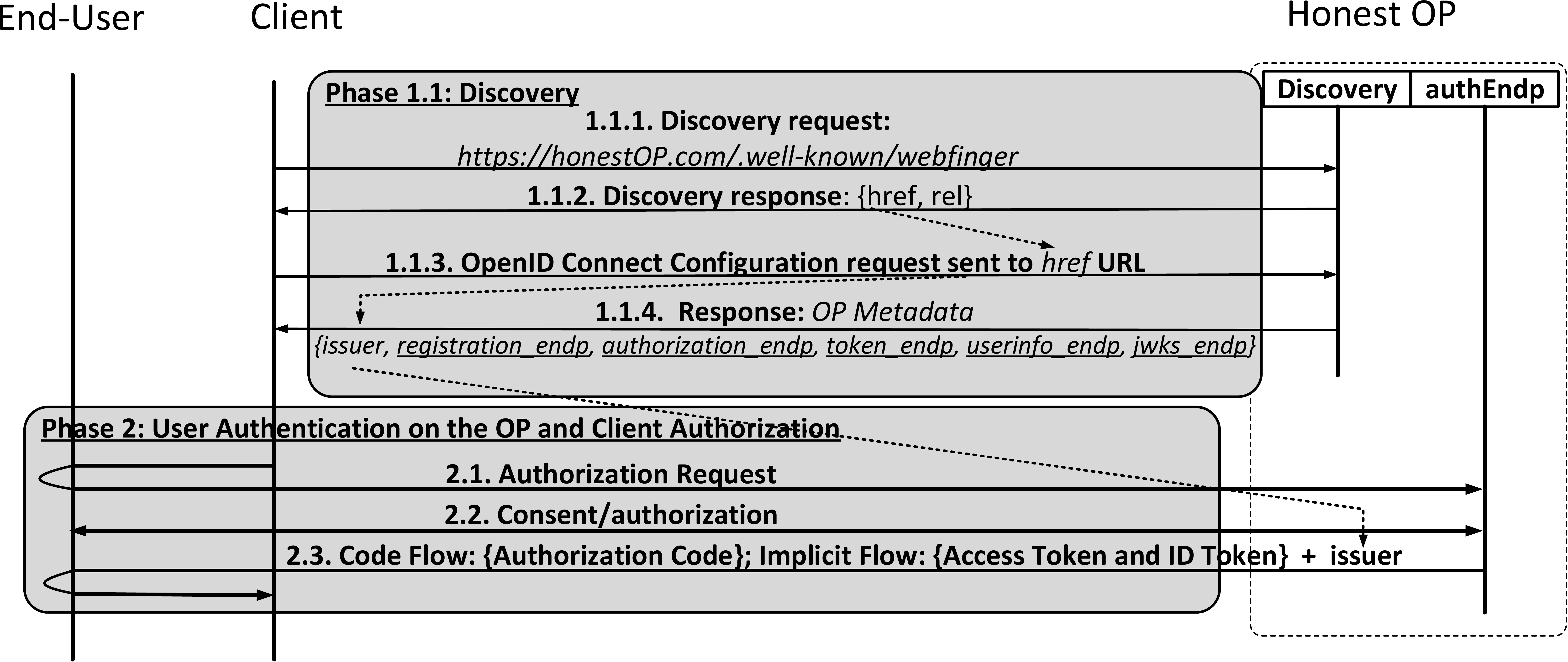}
	\caption{Binding Phase 1.1 (Discovery) and 2.3 (Authentication Response) via the \emph{issuer} element.}
	\label{fig:countermeasureBindDiscovery}
\end{figure}

\paragraph{Binding Discovery and Client Registration Phase}
We discussed our findings with the OAuth Working group.
As a result, the OAuth  specification will be extended in such a way that the Discovery phase is bound to the Authentication Response sent in Phase 2.
We depicted this approach in \autoref{fig:countermeasureBindDiscovery}.

The protocol flow is the same to its current specification with only one small change:
In Step 2.3, we added the \emph{issuer} value as an additional parameter.
Now, the Client is able to detect the attack since the issuer in the Discovery document is \emph{http://malicious.com} (cf. \autoref{lst:maliciousEndpoints}), but the honest \gls{op} responds with its value \emph{https://honestOP.com}.

This countermeasure requires a very small extension of the current protocol flow and additional checks on the \gls{op}, but in comparison to the previous approaches, it offers the full flexibility of the Discovery and Dynamic Client Registration extensions.

This countermeasure does not prevent \gls{dos} and \gls{ssrf} attacks and should be considered by the implementation of the Client.

\paragraph{Summary}
By implementing CSRF and Clickjacking countermeasures on the Client the attack described in \autoref{sec:brokenenduserauth} will be mitigated. 
However, we believe that more general and protocol-based solution should be provided by the \gls{oidc} specification.
Thus, we prefer the countermeasure binding the Discovery and Client Registration Phase.
Additionally, we advise the usage of both flows \emph{client\_secret\_jwt} or \emph{private\_key\_jwt} for Client authentication avoiding the transmission of the \emph{client\_secret} between the Client and the \gls{op}.

In order to reduce the impact of \gls{dos} attacks, the Client should expect short messages. Thus, it can be configured to wait small period for an answer and accepts messages less than several KBytes.
The Client should restrict the usage of URLs, protocols and Ports in order to reduce the attack surface of \gls{ssrf} attacks. For instance only HTTP requests with destination port 443 or 80 should be allowed.

%% file: sections/relatedWork.tex
\section{Related Work}
\label{sec:relatedwork}

\paragraph{Attacks on \gls{sso} systems}
In 2012, Wang et al.~\cite{microsoft} concentrated on real-life \gls{sso} and the analysis of \gls{sso} protocols and implementation flaws via Browser related messages (BRM).
The authors have well demonstrated the problems related to token verification with different attacks.
Additionally, they introduced a tool named BRM-Analyzer, which analyses the traffic passing through a user's browser and detects abnormalities in the protocol flow, attempting to notice attacks.
However, since the \gls{mep} attack described in this paper follows the protocol specification, no abnormalities can be detected by tools like BRM-Analyzer.
Moreover, the BRM-Analyzer cannot observe the communication during the Discovery and Dynamic Registration phase, since the communication between Client and \gls{op} occurs directly and not via the End-User's browser.

In 2012, Sun et al.~\cite{sun2012devil} analyzed nearly 100 \gls{oauth} implementations, and found serious security flaws in many of them.
The research focused on the impact of classical web attacks like \gls{xss}, \gls{csrf} and TLS misconfiguration on the \gls{oauth} implementations.
In~\cite{HomakovFacebookHackOAuth,HomakovOAuthQuestions,HomakovGitHubHackOAuth,GoldschlagerFacebookHackAgain,GoldschlagerFacebookHack,ssoscan}, further attacks on \gls{oauth} implementations were discovered and reported.
However, all these works concentrated on individual attacks and especially implementation misconfiguration.

In 2013, Wang et al. introduced a systematic process for identifying critical assumptions in SDKs, which led to the identification of exploits in constructed Apps resulting in changes in the \gls{oauth} specification~\cite{Wang:2013:ESU:2534766.2534801}.
Chen et al. revealed in 2014 serious vulnerabilities in \gls{oauth} applications on mobile devices caused by the developer's misinterpretation of the \gls{oauth} protocol~\cite{demystifiedOAuthCCS14}.

In 2014 Cao et. al. studied vulnerabilities in existing \gls{sso} protocols that allow impersonation attacks and analyzed the main reasons leading to these flaws~\cite{raey}. The authors concentrated only on phase 2 and phase 3 of the \gls{sso} protocol flow and did not consider phase 1. Interestingly, the authors of the paper recognized that one main problem in \gls{sso} is the lack of authentication between the \gls{op} and Client, which is part of the Malicious Endpoints attack.

Please note that none of the previous papers considers the \gls{oidcp} flow and the Discovery and Dynamic Registration phases.

\paragraph{Formal approaches}
An open problem that we see as important future work (see below) is the introduction of a formal language and an verifier tool that will be able to automatically detect vulnerabilities as described in this paper. 
In 2012 Sun et al.~\cite{journals/compsec/SunHB12} provided a semi-automated analysis on \gls{oid} by modeling \gls{csrf}, replay, impersonation, parameter forgery and session swapping attacks. 
However, the authors did not consider that any of the \gls{op}'s components can act maliciously (e.g. the Discovery service). 
In 2014 Fett et al.~\cite{FettKuestersSchmitz-SP-2014} introduced a formalization for the (now defunct) \gls{sso} service BrowserID. 
Nevertheless, in this formal model they still perform a {\em manual} security analysis. 

We refrain from modeling \gls{oidc} in such a formal model for two reasons: (1) We believe that a thorough understanding of (second-order) vulnerabilities is essential to developing a formal model for {\em automated} analysis, and therefore concentrate on readability. (2) The model from~\cite{FettKuestersSchmitz-SP-2014} depicts only BrowserID and thus has to be modified for each \gls{sso} protocol separately.

\paragraph{Second-order vulnerabilities}
In 2014 Dahse et. al. introduced an approach for static detection of second-order vulnerabilities in web applications~\cite{Dahse2014}. The authors considered attacks like \gls{sqlinjection} and \gls{xss} and methods to prevent such vulnerabilities.
More complex attacks including multiple steps for injecting attack vectors and their execution in distributed systems interacting with each other were not considered.
One year later Olivo et.al. introduced new class of \gls{dos} attacks referred to the second-order vulnerabilities~\cite{Olivo2015}. Additionally, the authors developed a static analysis approach for detecting second-order \gls{dos}  vulnerabilities in  web applications. 
More complex systems, for example distributed systems like \gls{sso}, were not considered and analyzed.

%% file: sections/conclusion.tex
\section{Discussion and Future Work}
\label{sec:conclusion}

In this paper, we analyzed the \gls{oidc} protocol considering all phases of the protocol.
During analyzing the \gls{oidc}'s the Discovery and Dynamic Registration phases we found several novel second-order vulnerabilities resulting in broken user authentication, \gls{dos}, \gls{ssrf} and injection attacks.
Summarized, we found an existing gap in previous security evaluations, since these concentrated only on the security critical phases -- \emph{Phase 2} and \emph{Phase 3}.

Speaking of \gls{sso}, other protocols like \gls{oid}~\cite[Section 7.3]{openid20}, SAML~\cite{samlDiscovery} and BrowserID~\cite{browserIDDiscovery} support features similar to the Discovery and Dynamic Registration extensions described in \gls{oidc}.
It is essential that these protocols are further studied to avoid similar security gaps.
We refer this research to the future work.

Second-order vulnerabilities in distributed systems like \gls{sso} are barely studied.
The explanation for this gap is the complexity of such distributed systems.
Since this complexity has not yet been tamed by an automated analysis tool, many of the \gls{sso} vulnerabilities today are discovered by manual security evaluation, which is time consuming and inefficient.

Developing such an automated analysis tool requires deep knowledge of the distributed information flows in a \gls{sso} system, and potential vulnerabilities. 
In this paper, we aim to provide exactly this information for the novel and important \gls{oidc} \gls{sso} protocol, especially for the discovery phase.

An important task in the future is the development of techniques and automated tools facilitating the modeling, evaluation and detection of security issues in distributed systems like \gls{sso}.
Currently, there is a gap, which should be in the scope of further researches.